\documentclass[pra,twocolumn,showpacs]{revtex4}

\usepackage[cp866]{inputenc}
\usepackage[english]{babel}
\usepackage{amsmath}
\usepackage{dcolumn}
\usepackage{epsfig}

\newcommand{\inputEps}[3]{
            \centerline{\epsfxsize=#1 \epsfbox{#3} }
\vskip 12pt {\center\small{#2}} \vskip 18pt}

\begin{document}

\title{\bf Two-color interference stabilization of atoms}
\author{M.V. Fedorov}
\email{fedorov@ran.gpi.ru}
\author{N.P. Poluektov}
\email{nickel@aha.ru}
\affiliation{
General Physics Institute, Russian Academy of Science}

\date{\today}

\begin{abstract}

The effect of interference stabilization is shown to exist in a
system of two atomic levels coupled by a strong two-color laser
field, the two frequencies of which are close to a two-photon
Raman-type resonance between the chosen levels, with open channels
of one-photon ionization from both of them. We suggest an
experiment, in which a rather significant (up to 90 $\%$)
suppression of ionization can take place and which demonstrates
explicitly the interference origin of stabilization. Specific
calculations are made for H and He atoms and optimal parameters of
a two-color field are found. The physics of the effect and its
relation with such well-known phenomena as LICS and population
trapping in a three-level system are discussed.

\end{abstract}

\maketitle

\section{Introduction}
\subsection{Interference stabilization}

Interference stabilization of Rydberg atoms, or strong-field
suppression of photoionization, is known \cite{Mov}, \cite{Book}
to be a phenomenon related to the coherent re-population  of
levels neighboring to the initially populated one. Such a
re-population arises owing to Raman-type transitions via the
continuum and in the case of a single-color field it can be
efficient only if the field is strong enough. Specifically, the
strong-field criterion for the effect of interference
stabilization is formulated qualitatively as the requirement for
the ionization width $\Gamma_i^{(n)}$ of the initially populated
atomic level $E_n$ to be larger than the spacing between
neighboring levels,
\begin{equation}
 \label{threshold}
 \Gamma_i^{(n)}>|E_n-E_{n\pm 1}|,
\end{equation}
where $n$ is the principal quantum number. The ionization width is
determined here as the rate of ionization calculated with the help
of the Fermi Golden Rule. However, repopulation of neighboring
Rydberg levels is provided, actually, by the off-diagonal terms of
the tensor of ionization widths $\Gamma_i^{(n',n)}$. In the
approximation of adiabatic elimination of the continuum (which
includes, in particular, the well-know rotating wave
approximation, \cite{Book}) this tensor is determined as a direct
generalization of the Fermi Golden Rule expression for
$\Gamma_i^{(n)}$
\begin{equation}
 \label{tensor}
 \begin{array}{c}
 \Gamma_i^{(n',n)}=\frac{\pi}{2}\;\varepsilon_0^2\;\langle n'|d|E\rangle
 \langle E|d|n\rangle\Big|_{E=E_n+\omega},
 \end{array}
\end{equation}
where $\varepsilon_0$ and $\omega$ are the laser field-strength
amplitude and frequency, $d$ is the projection of the atomic
dipole moment upon the direction of light polarization, $E$ is the
energy of the atomic electron in the continuum, and atomic units
are used throughout the paper if not indicated differently. So,
the next crucial assumption in the theory of interference
stabilization is that all the components of the tensor
(\ref{tensor}) are approximately equal to each other
\begin{equation}
 \label{equal}
 \Gamma_i^{(n',n)}\approx\Gamma.
\end{equation}
This assumption is pretty well fulfilled for high atomic Rydberg
levels, $n, n'\gg 1$, $|n-n'|\ll n$ (see explanations in
\cite{Book}). It should be noted also that for Rydberg levels
their ac Stark shift, as well as the shift of the ionization
threshold are equal approximately to the ponderomotive energy
$\varepsilon_0^2/4\omega^2$ and identical to each other, and this
common shift does not affect either the dynamics of
photoionization from Rydberg levels or the effect of interference
stabilization.

The simplest model, in which the effect of interference
stabilization exists, is the model of two close atomic levels
$E_1$ and $E_2$ connected with each other by the Raman-type
transitions via the continuum, for which the conditions
(\ref{threshold}) and (\ref{equal}) are fulfilled and the ac Stark
shift has the same features as described above for Rydberg levels
and, actually, can be ignored.

Both in two-level and multilevel systems there are several
different theoretical approaches one can use to solve the problems
of strong-field photoionization and stabilization. One of them is
based on the use of quasienergy or "dressed-state" analysis. The
total wave function of an atomic electron in a light field can be
expanded in a series of the field-free atomic eigenfunctions
\begin{equation}
 \label{expansion}
 \Psi=\sum_n C_n(t)\,\psi_n + {\rm continuum}.
\end{equation}
In the approximation of adiabatic elimination of the continuum
equations for the coefficients $C_n(t)$ are stationary, and in the
simplest case of the two-level system they have the form
\begin{equation}
 \label{2-lev-eq}
 \begin{array}{c}
 i\dot{C}_1(t)-E_1C_1(t)=-\,\frac{i}{2}\,\Gamma\,\left[C_1(t)+C_2(t)\right]\\
 \,\\
 i\dot{C}_1(t)-E_2C_1(t)=-\,\frac{i}{2}\,\Gamma\,\left[C_1(t)+C_2(t)\right],
 \end{array}
\end{equation}
where the approximation (\ref{equal}) is assumed to be fulfilled.

As equations (\ref{2-lev-eq}) are stationary, they have solutions
of the form $C_{1,2}\propto\exp(-i\gamma\,t)$, where $\gamma$ is a
complex quasienergy. When this exponential dependence on $t$ is
substituted into Eqs. (\ref{2-lev-eq}), they turn into a set of
two algebraic homogeneous equations, which has a nonzero solution
if its determinant turns zero. This is the condition from which
the two quasienergies of the field-driven two-level system have to
be found, and the result is given by
\begin{equation}
 \label{2-lev-qe}
 \begin{array}{c}
 \gamma_\pm=\frac{1}{2}\left\{E_1+E_2-i\,\Gamma\pm
 \sqrt{(E_2-E_1)^2-\Gamma^2}\right\}.
 \end{array}
\end{equation}
From here we see that, indeed, a drastic change in the form of the
solutions occurs when the interaction constant $\Gamma$ becomes
larger than the level spacing $E_2-E_1$. The point $\Gamma=
E_2-E_1$ is the branching point, below which (at $\Gamma<E_2-E_1$)
the root square is real in Eq. (\ref{2-lev-qe}), whereas above the
branching point (at $\Gamma>E_2-E_1$) it becomes imaginary. The
imaginary parts of the quasienergies $\gamma_\pm$ (\ref{2-lev-qe})
are shown in Fig. 1, and they determine the field-dependent widths
of the two quasienergy levels $\Gamma_\pm(\varepsilon_0)\equiv
2|{\rm Im}[\gamma_\pm(\Gamma)]|$, where
$\Gamma\propto\varepsilon_0^2$ (\ref{tensor}). One of the two
branches arising at $\Gamma>E_2-E_1$ ($\gamma_+(\Gamma)$)
corresponds to a narrowing quasienergy level whose width
$\Gamma_+(\varepsilon_0)$ falls with a growing field-strength
amplitude. This corresponds to an increasing life-time of this
quasienergy level and to stabilization of an atomic population at
this level.

\vskip 12pt

\inputEps{180pt}{Fig. 1. The functions ${\rm Im}[\gamma_\pm(\Gamma)]$ (\ref{2-lev-qe}).}
{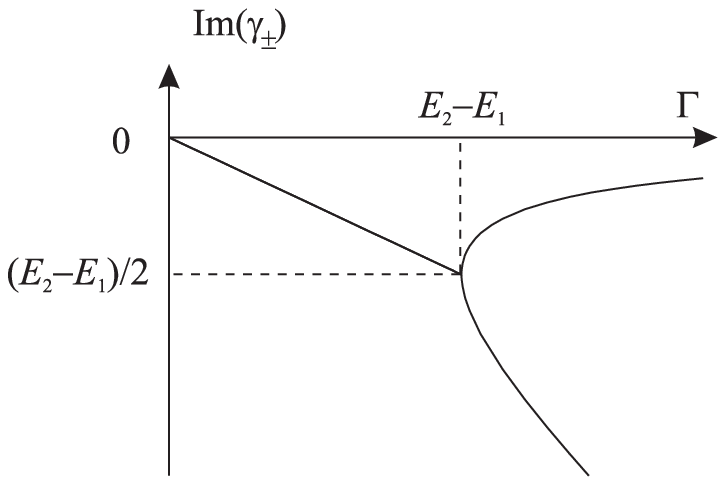}

\subsection{Laser-Induced Continuum Structures, Autoionizing Resonances,
Dark States, and Population Trapping}

Though rather attractive by its simplicity, an isolated two-level
system obeying the requirement (\ref{equal}) hardly can be easily
found in the usual atomic spectra. This is the reason why here we
consider another scheme, in which two atomic levels with
significantly different energies and bound-free dipole matrix are
connected with each other by Raman-type transitions via the
continuum in a two-color field (Fig. 2). Such a scheme has been
widely discussed in literature \cite{Armstr, Heller, Andr, Lambr,
PQE, Knight, Mag, Hutch, Shao, Caval, Half} in connection with the
phenomenon of Light-Induced Continuum Structure (LICS), briefly
outlined below. The process we suggest and investigate, as well as
its similarity and differences with LICS are discussed in
Subsection ${\bf{\rm C}}$.

In LICS, one of the two fields of a two-color light  is assumed to
be strong [($\omega_2$, $\varepsilon_2$), the pump] and the other
one - weak [($\omega_1$, $\varepsilon_1$), the probe], where
$\omega_{1,\,2}$ and $\varepsilon_{1\,2}$ are the corresponding
frequencies and field-strength amplitudes. In a scheme of Fig. 2
$\Delta$ is the Raman-type resonance detuning.
\begin{equation}
 \label{delta}
 \Delta=E_2+\omega_2-E_1-\omega_1
\end{equation}

\vskip 10pt
\inputEps{160pt}{Fig. 2. A scheme of two atomic levels under the
conditions of a Raman-type resonance in a two-color filed.}
{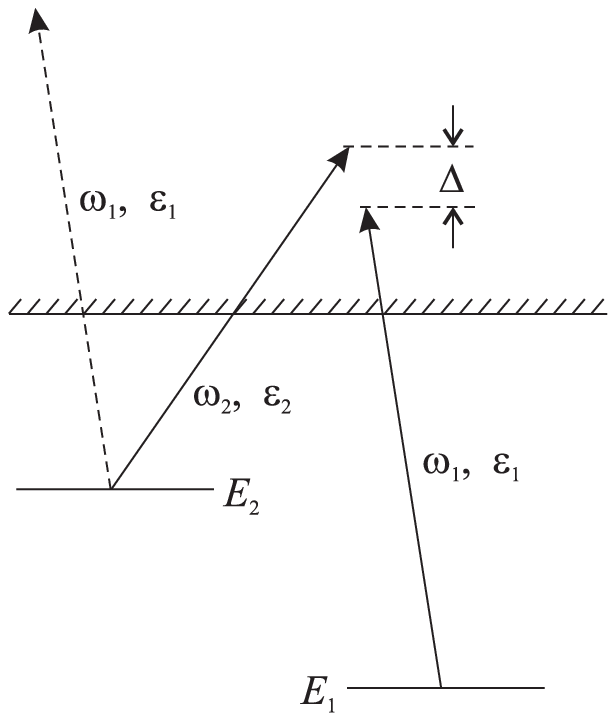}

Under the action of the pump the first Floquet satellite of the
level $E_2$ takes the form of an autoionizing-like level
$E_2+\omega_2$ at the background of the continuum with the width
equal to the ionization width of the level $E_2$,
$\Gamma_i^{(2)}\equiv\Gamma_i^{(2,2)}$ (\ref{tensor}) with
$\varepsilon_0$ substituted by $\varepsilon_2$. If all the
population is concentrated initially at the level $E_1$, the probe
field ionizes the atom and takes the Floquet satellite
$E_2+\omega_2$ for an almost real autoionizing level, which gives
rise to the typical asymmetric Fano-profile-like structure of the
dispersion curve $w_i(\omega_1)$ (Fig. 3).

\vskip 12pt

\inputEps{160pt}{Fig. 3. The Fano profile of the dispersion curve
$w_i(\omega_1)$ at an autoionizing-like resonance.} {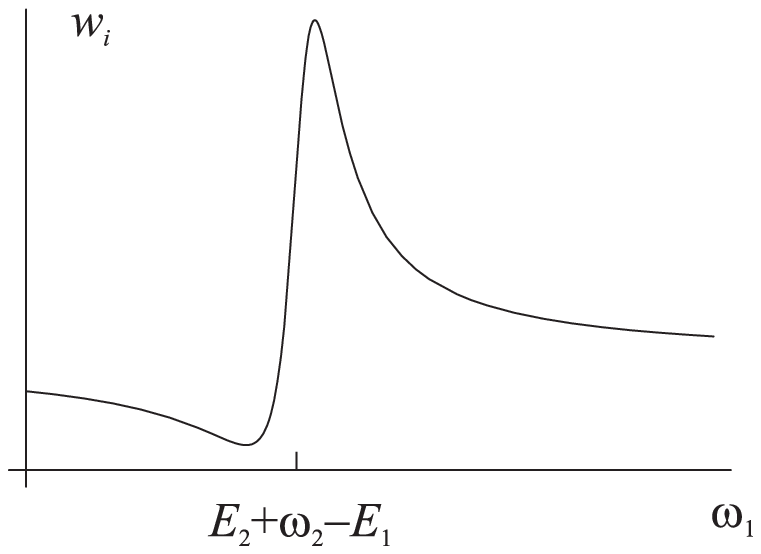}

\noindent The Fano minimum  of the curve $w_i(\omega_1)$ arises
owing to interference of direct and indirect transitions to the
continuum ($E_1\rightarrow E$ and $E_1\rightarrow E'\rightarrow
E_2\rightarrow E$). But this is not yet a stabilization understood
as an increasing suppression of ionization with a growing light
intensity. An experiment we suggest and discuss below can
demonstrate explicitly such an interference suppression and
stabilization of an atom in its bound states. We will assume that
both fields can be equally strong and in this sense the effect we
consider can be referred to as a strong-field LICS.

It should be noted that the channel of ionization shown in Fig. 2
by a dashed line (taken into  account for the first time in Ref.
\cite{Andr}) determines a nonzero height of the Fano curve in its
minimum. As we will see below, this is just the competition of
this non-interfering channel of ionization with interfering ones
that is crucially important for optimization of stabilization in
its dependence on light intensities.

At last, another well-known analogue of LICS is related to single-
and double strong-field resonances at real autoionizing atomic
levels \cite{PQE,Eberly, Lambr2, AKF, Rahman}. The physics of this
phenomenon and LICS are close though there are evident differences
concerning mechanisms of level broadening. These differences make
autoionizing resonances not as closely related to the phenomenon
under consideration as LICS. The same can be said about dark
states and population trapping in a three-level system
\cite{Gozzini, Ari-Orri, Stroud, Arimondo}. The physics of all
these phenomena is alike though important details are different.
In particular, this concerns intensity-dependent mechanisms of
level broadening and an important role of the noninterfering
ionization channel specific for the scheme under consideration and
missing in a three-level scheme. Besides, the atomic continuous
spectrum is so much wider than any discrete third level that this
makes characteristic intensities in the phenomenon to be discussed
absolutely different from those in the population trapping effect.

\subsection{An experiment we suggest}

Let us assume that the initially populated level in a scheme of
Fig. 2 is $E_1$. At the first stage, by considering ionization of
an atom by the field $\varepsilon_1$ alone (with
$\varepsilon_2=0$), we select the peak light intensity
$I_1=\varepsilon_1^2/8\pi$ and pulse duration $\tau$ high and long
enough to provide almost complete ionization of an atom by a
pulse, $w_i(I_1,\,I_2=0)=1$. Then, by adding the field
$\varepsilon_2$ we expect that under proper conditions, owing to
interference effect, the combined action of two fields will result
in a significant suppression of ionization. In other words, we
expect that the function $w_i(I_1,\,I_2)$ in its dependence on
$I_2$ at a given selected value of $I_1$ will start from 1 at
$I_2=0$, then it will have a minimum at some intensity
$I_2=I_{2\,0}\sim I_1$, and then, at higher intensity $I_2$,
$w_i(I_1,\,I_2)$ will return to one again. The region around
$I_{2\,0}$ will be interpreted as a stabilization window. The
interference origin of stabilization is evident, because with a
growing $I_2$ (at a given $I_1$), we increase an energy that can
be put into an atom. But, counterintuitively, this additional
energy results in a slower rather than faster ionization, and this
can be explained only by interference effects.

We will find values of the resonance detuning $\Delta$ and the
ratio of intensities $I_2/I_1$ which optimize the stabilization
effect.

\section{The main equations}
Compared to a simplified  system of two close levels in a
single-color field described in the Introduction, to characterize
appropriately a system of Fig. 2 in a two-color arbitrary strong
field, in addition to ionization broadening and mixing of levels,
we have to take into account also their shifts and mixing arising
owing to the ac Stark effect. Both these effects can be described
in terms of the complex polarizability tensor $\alpha_{i,j}$, $i,
j=1,2$,
$$
 \alpha_{i,i}(\omega)\equiv\alpha_i(\omega)
$$
\begin{equation}
 \label{i,i}
 =\int dE\;|d_{i\,E}|^2\left(\frac{1}{E-E_i-\omega-i\delta}
 +\frac{1}{E-E_i+\omega}\right)
\end{equation}
and
$$
 \alpha_{2\,1} =\int
dE\;d_{2\,E}d_{E\,1}\left(\frac{1}{E-E_1-\omega_1-i\delta}\right.
$$
\begin{equation}
 \label{off-diag}
 +\left.\frac{1}{E-E_1+\omega_2}\right)\approx\alpha_{12},
\end{equation}
where integration over $E$ includes summation over intermediate
discrete states.

For the two-color scheme (Fig. 2), similarly to (\ref{expansion}),
in the rotating wave approximation the wave function of an atomic
electron can be written as
\begin{equation}
 \label{exp2}
 \Psi=C_1(t)\,e^{i\omega_1 t}\psi_1+C_2(t)\,e^{i\omega_2 t}\psi_2+
 {\rm continuum}.
\end{equation}
As well as Eqs. (\ref{2-lev-eq}), equations for the probability
amplitudes $C_1(t)$ and $C_2(t)$  are obtained from the Schr${\rm
{\ddot o}}$dinger equation with the help of the procedure of
adiabatic elimination of the continuum. In terms of $\alpha_{i,j}$
(\ref{i,i}), (\ref{off-diag}), these equations can be presented in
the form
\begin{equation}
 \label{2-lev-eq-gen}
 \begin{array}{c}
 i{\dot C}_1-(\widetilde{E}_1(t)+\omega_1)C_1=
-\frac{1}{4}\varepsilon_{1\,0}(t)\varepsilon_{2\,0}(t)\alpha_{12}\,C_2,\\
 \,\\
 i{\dot C}_2-(\widetilde{E}_2(t)+\omega_2)C_2
=-\frac{1}{4}\varepsilon_{1\,0}(t)\varepsilon_{2\,0}(t)\alpha_{21}\,C_1,
 \end{array}
\end{equation}
where $\widetilde{E}_i(t)$ are the slowly time-dependent adiabatic
complex energies of the ac-Stark-shifted and broadened levels
\begin{equation}
 \label{dressed-levels}
 \begin{array}{c}
 \widetilde{E}_i(t)
 =E_i-\frac{1}{4}\Bigl\{\alpha_i(\omega_1)\varepsilon_{1\,0}^2(t)
 +\alpha_i(\omega_2)\varepsilon_{2\,0}^2(t)\Bigr\}.
 \end{array}
\end{equation}

\section{Quasienergies}

For the time-dependent pulse envelopes $\varepsilon_{1\,0}(t)$ and
$\varepsilon_{2\,0}(t)$, Eqs. (\ref{2-lev-eq-gen}) have to be
solved as the initial-value problem. In the model of constant
field-strength amplitudes these equations have stationary
solutions $C_{1,2}\propto\exp(-i\gamma t)$, where, as previously,
$\gamma$ is the complex quasienergy for which we get the solutions
generalizing those of Eq. (\ref{2-lev-qe})
\begin{equation}
 \label{2-lev-qe-gen}
 \begin{array}{c}
 \gamma_\pm=\frac{1}{2}\Big\{\widetilde{E}_1+\omega_1+
 \widetilde{E}_2+\omega_2\pm D\Big\},
 \end{array}
\end{equation}
where
\begin{equation}
 \label{D}
 D=\sqrt{\widetilde{\Delta}^2+\frac{1}{4}\alpha_{12}\alpha_{21}
 \varepsilon_{1\,0}^2\varepsilon_{2\,0}^2}
\end{equation}
and $\widetilde{\Delta}$ is the time-dependent complex detuning
for the ac-Stark-shifted and broadened levels
(\ref{dressed-levels})
\begin{equation}
 \label{dressed-detun}
 \widetilde{\Delta}=\widetilde{E}_2+\omega_2-\widetilde{E}_1-\omega_1.
\end{equation}

Imaginary parts of the energies $\widetilde{E}_i$
(\ref{dressed-levels}) are related to the ionization widths
determined by imaginary parts ${\rm
Im}\left(\alpha_i\right)\equiv\alpha_i^{\prime\prime}$ of the
polarizabilities $\alpha_i$ (\ref{i,i})
\begin{equation}
 \label{widths}
 \begin{array}{c}
 {\rm Im}\left( E_1\right)=-\frac{1}{2}\Gamma_1,\\
 \,\\
 {\rm Im}\left(
 E_2\right)=-\frac{1}{2}\left(\Gamma_2^{(1)}+\Gamma_2^{(2)}\right),
 \end{array}
\end{equation}
where
\begin{equation}
 \label{widths-pol}
 \begin{array}{c}
 \Gamma_1=\frac{1}{2}\alpha_1^{\prime\prime}(\omega_1)\varepsilon_{1\,0}^2,\\
 \,\\
 \Gamma_2^{(1)}=\frac{1}{2}\alpha_2^{\prime\prime}(\omega_1)\varepsilon_{1\,0}^2,\;\;
 {\rm and}\;\;
 \Gamma_2^{(2)}=\frac{1}{2}\alpha_2^{\prime\prime}(\omega_2)\varepsilon_{2\,0}^2.
 \end{array}
 \end{equation}
 The width $\Gamma_2^{(1)}$ is determined by transitions from the
 level $E_2$ under the action of the field $\varepsilon_{1\,0}$ (the dashed line in Fig.
 2). As $E_2>E_1$ and $\omega_1>\omega_2$, typically,
\begin{equation}
 \label{smaller}
 \alpha_2^{\prime\prime}(\omega_1)\ll\alpha_1^{\prime\prime}(\omega_1)\;\;{\rm
 or}\;\; \Gamma_2^{(1)}\ll\Gamma_1.
\end{equation}
Similarly to (\ref{widths-pol}), the off-diagonal component of the
polarizability tensor (\ref{off-diag}) determines the off-diagonal
component of the ionization-width tensor
\begin{equation}
 \label{widths-pol-off-diag}
 \begin{array}{c}
  \Gamma_{12}=\frac{1}{2}\alpha_{12}^{\prime\prime}\varepsilon_{1\,0}\varepsilon_{2\,0}
  =\sqrt{\Gamma_1\Gamma_2^{(2)}},
 \end{array}
\end{equation}
which assumes, in particular, that
$\alpha_{12}''=\sqrt{\alpha_1''(\omega_1)\,\alpha_2''(\omega_2)}$.

Imaginary parts of quasienergies $\gamma_\pm$ (\ref{2-lev-qe-gen})
are related to the width of quasienergy levels $\Gamma_\pm$
\begin{equation}
 \label{gamma-Gamma}
 \Gamma_\pm=-2{\rm Im}\left(\gamma_\pm\right).
\end{equation}

\section{Probability of ionization}

The described above quasienergy solutions of Eqs.
(\ref{2-lev-eq-gen}) are most appropriate for solving the
initial-value problem in the case of pulses with rectangular
envelopes $\varepsilon_{1,2\;0}(t)$, with sudden turn-on and
turn-off (at $t=0$ and $t=\tau$) and
$\varepsilon_{1,2\;0}(t)=const$ at $0<t<\tau$. With known
quasienergies $\gamma_\pm$ (\ref{2-lev-qe-gen}) the time-dependent
probability amplitudes $C_{1,2}(t)$ to find an atom in its bound
states $\psi_1$ and $\psi_2$ can be presented in the form
\begin{equation}
 \label{in-val-sol}
 C_{1,2}(t)=A_{1,2}^+\exp(-i\,\gamma_+\,t)+A_{1,2}^-\exp(-i\,\gamma_-\,t),
\end{equation}
where $A_{1,2}^\pm$ are constants to be found from the initial
conditions
\begin{equation}
 \label{in-cond}
 A_1^++A_1^-=1\;\;\;{\rm and}\;\;\;A_2^++A_2^-=0
\end{equation}
and from equations connecting $A_1^\pm$ with $A_2^\pm$. The latter
follow, e.g., from the first of Eqs. (\ref{2-lev-eq-gen})
\begin{equation}
 \label{dop-eq}
 \begin{array}{c}
 \gamma_\pm A_1^\pm-(\widetilde{E}_1+\omega_1)A_1^\pm=
 -\frac{1}{4}\varepsilon_{1\,0}\varepsilon_{2\,0}\alpha_{12}\,A_2^\pm.
 \end{array}
\end{equation}

The total residual probability $w_{res}(\tau)$ to find an atom in
bound states at $t=\tau$ is given by the sum of partial
probabilities $w_1(\tau)$ and $w_2(\tau)$ to find an atom at
levels $E_1$ and $E_2$,
\begin{equation}
 \label{w-res}
 w_{res}(\tau)=w_1(\tau)+w_2(\tau),
\end{equation}
where
\begin{equation}
 \label{w-12}
 w_{1,\,2}(\tau)=|C_{1,\,2}(\tau)|^2
 =\Big|A_{1,\,2}^+ e^{-i\,\gamma_+\,\tau}+A_{1,\,2}^-
 e^{-i\,\gamma_-\,\tau}\Big|^2.
\end{equation}
The probability of ionization is given by
$w_i(\tau)=1-w_{res}(\tau)$. Eqs. (\ref{in-cond}) and
(\ref{dop-eq}) are easily solved to give, explicitly,
\begin{equation}
 \label{w-1-expl}
 w_1(\tau)=\frac{1}{4}\left|\left(1-\frac{\widetilde{\Delta}}{D}\right)e^{-i\gamma_+\tau}+
 \left(1+\frac{\widetilde{\Delta}}{D}\right)e^{-i\gamma_-\tau}\right|^2
\end{equation}
and
\begin{equation}
 \label{w-2-expl}
 w_2(\tau)=\frac{\alpha_{12}^2
 \varepsilon_{1\,0}^2\varepsilon_{2\,0}^2}{16\,D^2}\left|e^{-i\gamma_+\tau}-
 e^{-i\gamma_-\tau}\right|^2.
\end{equation}

 In the case of pulses with smooth envelopes
 $\varepsilon_{1,\,2}(t)$ quasienergy solutions are not so useful
 for solving the initial-value problem, and one has to solve directly
 Eqs. (\ref{2-lev-eq-gen}) for the time-dependent probability
 amplitudes $C_{1,\,2}(t)$.

\section{Scaling effect and relative units}

With the help of a phase transformation
\begin{equation}
 \label{phase}
 C_i(t)=\exp\left\{-i(E_1+\omega)t\right\}A_i(t)
\end{equation}
equations (\ref{2-lev-eq-gen}) can be reduced to an asymmetric
form
\begin{equation}
 \label{eq-assym}
 \begin{array}{lcr}
 &i{\dot A}_1+\frac{1}{4}\Bigl\{\alpha_1(\omega_1)\varepsilon_{1\,0}^2(t)
 +\alpha_1(\omega_2)\varepsilon_{2\,0}^2(t)\Bigr\}A_1&\\
 &=-\frac{1}{4}\varepsilon_{1\,0}(t)\varepsilon_{2\,0}(t)\alpha_{12}\,A_2&\\
 {\rm and}&&\\
 &i{\dot A}_2-\Big(\Delta -\frac{1}{4}\Bigl\{\alpha_2(\omega_1)\varepsilon_{1\,0}^2(t)
 +\alpha_2(\omega_2)\varepsilon_{2\,0}^2(t)\Bigr\}\Big)A_2&\\
 &=-\frac{1}{4}\varepsilon_{1\,0}(t)\varepsilon_{2\,0}(t)\alpha_{21}\,A_1,&
 \end{array}
\end{equation}
where, as previously, $\Delta$ is the weak-field
two-photon-resonance detuning (\ref{delta}).

Though not as nice as (\ref{2-lev-eq-gen}), Eqs. (\ref{eq-assym})
are more convenient to describe the scaling effect existing in the
system under consideration. Let us assume that both pulse
envelopes $\varepsilon_{1\,0}(t)$ and $\varepsilon_{2\,0}(t)$
depend on time $t$ only via the ratio $t/\tau$, where $\tau$ is
the pulse duration common for both high- and low-frequency pulses.
Then, evidently, the arguments of the functions
$\varepsilon_{1\,0}(t)$ and $\varepsilon_{2\,0}(t)$ do not change
if we divide both $t$ and $\tau$ by the same factor $\lambda$,
$t\rightarrow t/\lambda$ and $\tau\rightarrow\tau/\lambda$.
Moreover, one can see easily that Eqs. (\ref{eq-assym}) do not
change too if we multiply simultaneously both low- and
high-frequency pulse peak intensities
$I_{1,2}=c\varepsilon_{1,2\,0}^2/8\pi$ and the weak-field detuning
$\Delta$ by the same factor $\lambda$. So, the solutions of Eqs.
(\ref{eq-assym}) are invariant with respect to the scaling
transformation:
\begin{equation}
 \label{scaling}
 \Delta\rightarrow\lambda\Delta,\;
 \varepsilon_{1,2\,0}^2\rightarrow\lambda\varepsilon_{1,2\,0}^2,\;
 \tau\rightarrow\tau/\lambda,\;t\rightarrow t/\lambda.
\end{equation}
with an arbitrary $\lambda$. This scaling effect can be important
for practice: parameters of an assumed experiment can be varied to
choose the most convenient conditions for observation the
two-color stabilization effect discussed in this paper. In
particular, by making laser pulses longer, one can use rather
moderate-intensity lasers, as it's shown below.

Owing to the the described scaling effect, it is convenient to
introduce and use the dimensionless ratio of intensities $x$,
interaction time $\theta$ and detuning $\delta$,
\begin{equation}
 \label{dim-less-x-t-d}
 x=\frac{I_2}{I_1},\;\theta=\tau\cdot I_1,\;\delta=\frac{\Delta}{I_1},
\end{equation}
dimensionless complex quasienergies
\begin{equation}
 \label{dim-less-qe}
 y_\pm=\frac{\gamma_\pm-E_1-\omega_1}{I_1},
\end{equation}
detuning for the ac Srark shifted and broadened levels
(\ref{dressed-levels})
$$
 \widetilde{\delta}=\frac{\widetilde{\Delta}}{I_1}
$$
\begin{equation}
 \label{relative}
 =\delta-\frac{1}{4}\Big\{\alpha_2(\omega_1)+\alpha_1(\omega_1)+
 \big[\alpha_1(\omega_2)+\alpha_2(\omega_2)\big]x\Big\},
\end{equation}
and widths of the fully dressed quasienergy levels
$$
 g_\pm=\frac{\Gamma_\pm}{I_1}=-2{\rm Im}[y_\pm]
$$
\begin{equation}
 \label{width-dim-less}
 =\frac{1}{2}\big[\alpha_1''(\omega_1)+\alpha_1''(\omega_2)x\big]-\widetilde{\delta}''
 \mp {\rm Im}\left(\sqrt{{\widetilde\delta}^{\,\displaystyle^{\,2}}+
 \frac{1}{4}\alpha_{12}^2 x}\right),
\end {equation}
where $I_{1,\,2}$, $\Delta$, $\tau$, $\gamma_\pm$, and
$\Gamma_\pm$ are in atomic units. Defined in such a way,
quasienergies $y_\pm$ and widths $g_\pm$ depend only on two
parameters, $x$ and $\delta$, whereas the probability of
ionization $w_i$ and the residual probability to find an atom in
bound states $w_{res}$ (\ref{w-res})-(\ref{w-2-expl}) depend on
three parameters, $x$, $\delta$, and $\theta$.

\section{Pulse shape}

The concepts of quasienergies and quasienergy functions are very
fruitful for an analysis exploiting a model of a rectangular pulse
envelope. Such an analysis is useful for clarification of physics
of the phenomenon under consideration. However, more realistic
laser pulse shapes are characterized by smooth envelopes. To
investigate a sensitivity of the results to be derived on the
pulse shape and its smoothing, we will consider pulse envelopes
$\varepsilon_{1,\,2}(t)$ of the form
\begin{equation}
 \label{sin^2}
  \varepsilon_{1,\,2\;0}(t)=\varepsilon_{1,\,2\;0}
 \times
 \displaystyle\frac{(1+a)\,\sin^2
 \left[\pi\left(N(a)\displaystyle\frac{t}{\tau}+\frac{1}{2}\right)\right]}
 {1+a\,\sin^2\left[\pi\left
 (N(a)\displaystyle\frac{t}{\tau}+\frac{1}{2}\right)\right]},
\end{equation}
where $\varepsilon_{1,\,2\;0}=const.$, $-1/2N(a)\leq t/\tau\leq
1/2N(a)$, $N(a)$ is the normalization factor
\begin{equation}
 \label{nonumber}
 N(a)=\frac{(1+a)^2}{a^2}\left\{1-\frac{2+3a}{2(1+a)^{3/2}}\right\}
\end{equation}
such that
\begin{equation}
 \label{half-width}
 \int_{-\tau/2N}^{\tau/2N}\varepsilon_{1,\,2\;0}^2(t)\,dt=\tau\times\varepsilon_{1,\,2\;0}^2,
\end{equation}
and $a$ is a smoothing parameter. At $a\rightarrow\infty$,
$N(a)\rightarrow$1 and the envelopes $\varepsilon_{1,\,2\;0}(t)$
(\ref{sin^2}) turn into the rectangular ones. At $a=0$, $N(0)=3/8$
and $\varepsilon_{1,\,2\;0}(t)$ (\ref{sin^2}) takes the form of a
pure $\sin^2$ pulse envelope
\begin{equation}
 \label{pure-sin^2}
 \varepsilon_{1,\,2\;0}(t)=\varepsilon_{1,\,2\;0}\times\sin^2\left[\pi
 \left(\frac{3t}{8\tau}+\frac{1}{2}\right)\right].
\end{equation}

\noindent By definition, for the pulse envelopes of the form
(\ref{sin^2}), at all values of the smoothing factor $a$ the peak
values of the field strengths and areas under the curves
$\varepsilon_{1,\,2}^2(t)$ (field energy per unit cross section)
are kept constant and equal to $\varepsilon_{1,\,2\;0}$ and
$\varepsilon_{1,\,2\;0}^2\times\tau$, correspondingly (see Fig.
4).  Hence, at all $a$ the parameter $\tau$ can be interpreted as
the pulse duration determined by the condition (\ref{half-width}).
For all $a$ the dimensionless pulse duration is determined as
previously, $\theta=I_1\times\tau$ with both $I_1$ and $\tau$
taken in atomic units.

\vskip 12pt

\inputEps{220pt}{Fig. 4. Pulse envelopes (\ref{sin^2}) at three different
values of the smoothing factor $a$.}{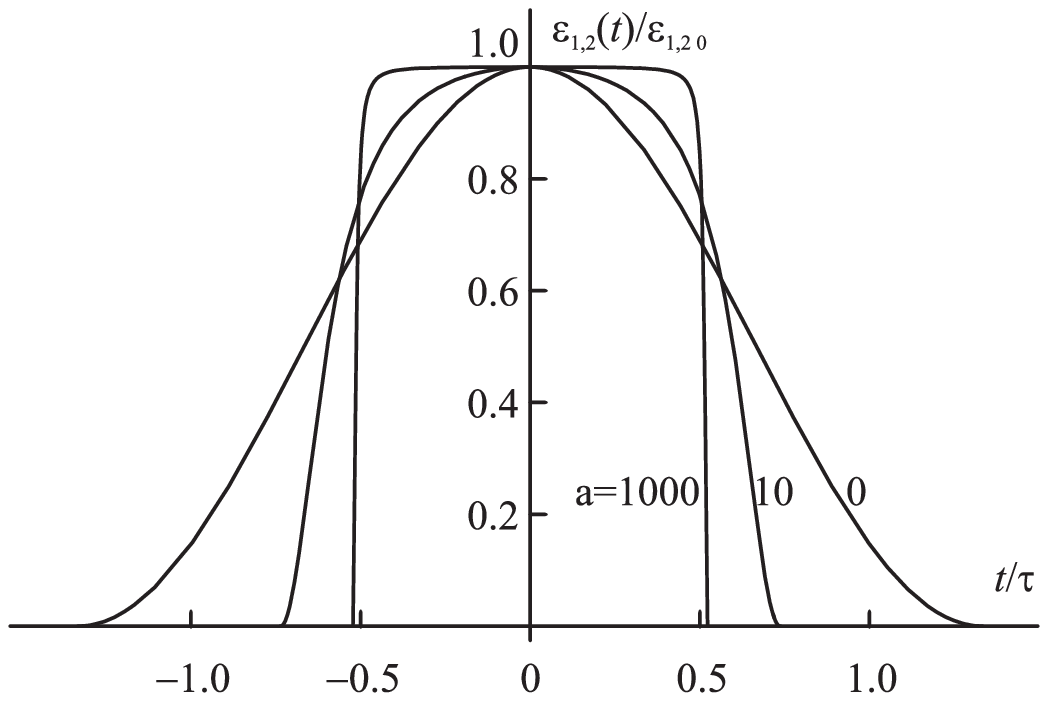}

\section{Optimization of the level-narrowing effect}

Optimization of the stabilization  and level-narrowing effects
assumes minimization of the width of the narrower quasienergy
level $g_+(\delta, x)$ with respect to two variables,  $\delta$
and $x$.  As a first step in a solution of this problem, let us
minimize $g_+(\delta, x)$ with respect to the field-free detuning
$\delta$ at a given $x$. Typically, at any given $x$, in
dependence on $\delta$, the functions $g_\pm(\delta, x=const)$
have well pronounced minimum and maximum (Fig. 5).

\vskip 12pt

\inputEps{180pt}{Fig. 5. The functions $g_\pm(\delta)\equiv
g_\pm(\delta, x=const)$(\ref{relative}).} {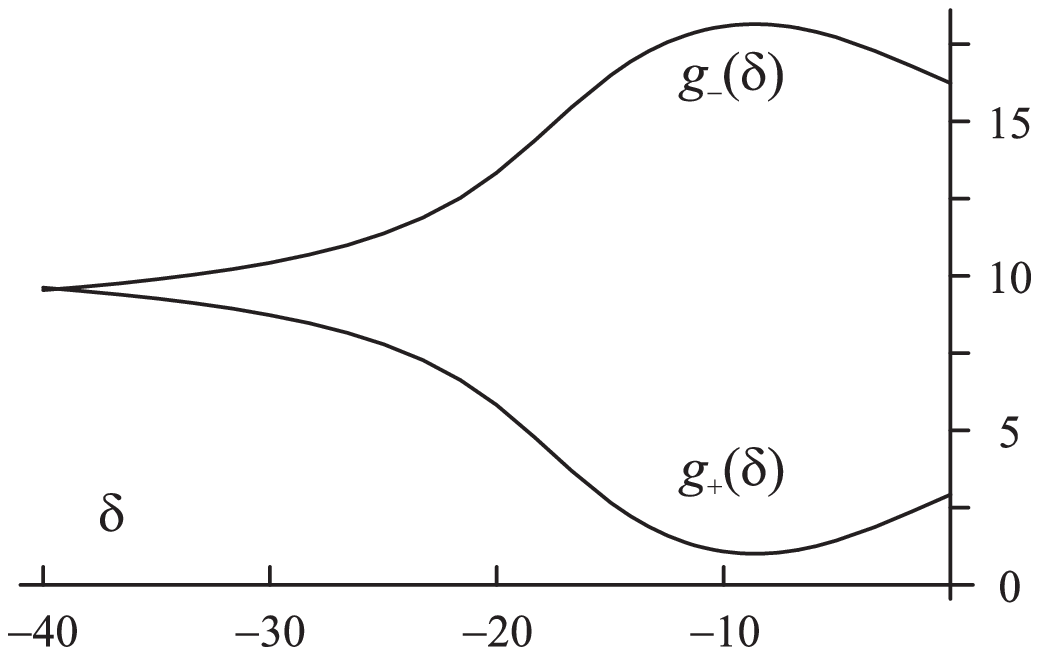}

\noindent Specifically, the curves of Fig. 5 are calculated for a
He atom at $x=0.1$. Details of these and many other calculations,
as well as the data about frequencies, atomic levels, and
polarizability tensors are given in the following Section. Here
the picture of Fig. 5 is shown as a typical example of the
dependencies $g_\pm(\delta, x=const)$. To find a position of the
extremum shown in Fig. 5, $\delta_{opt}(x)$, we have to solve the
equation $dg_\pm(\delta)/d\delta=0$. Direct calculations show that
this condition is satisfied if ${\rm
Im}\{[\widetilde{\delta}\cdot\alpha_{12}^*]^2\}=0$, which gives
two equations: ${\rm Re}[\widetilde{\delta}\cdot\alpha_{12}^*]=0$
and ${\rm Im}[\widetilde{\delta}\cdot\alpha_{12}^*]=0$. It can be
checked directly that only the second of these two equations
corresponds to the extremum we are looking for, and this equation
gives
\begin{equation}
 \label{extremum}
 \alpha_{12}'\,
 \widetilde{\delta}''-\,\alpha_{12}''\,\widetilde{\delta}'=0,
 \;\; \textrm{or}\;\;\widetilde{\delta}_{opt}=
 \frac{\alpha_{12}}{\alpha_{12}''}\,\widetilde{\delta\,}'',
\end{equation}
or
$$
 \delta_{opt}(x)=\frac{1}{4}\Big\{\alpha'_2(\omega_1)-\alpha'_1(\omega_1)-
 \frac{\alpha'_{12}}{\alpha''_{12}}[\alpha''_2(\omega_1)-\alpha''_1(\omega_1)]\Big\}
$$
\begin{equation}
\label{opt-detuning}
 +\frac{x}{4}\Big\{\alpha'_2(\omega_2)-\alpha'_1(\omega_2)-
 \frac{\alpha'_{12}}{\alpha''_{12}}[\alpha''_2(\omega_2)-\alpha''_1(\omega_2)]\Big\}.
\end{equation}
As explained above, $\delta_{opt}(x)$ is a value of the field-free
detuning, at which the width of one of the two quasienergy levels
($\gamma_+$) has a minimum with respect to $\delta$ at arbitrary
given $x$. As it's seen from Eq. (\ref{opt-detuning}) the optimal
detuning $\delta_{opt}(x)$ is a linear function of the ratio of
intensities $x=I_2/I_1$.

The second step in optimization conditions for stabilization
requires minimization of the width $g_+$ calculated at
$\delta=\delta_{opt}(x)$ with respect to the variable $x$. With
the help of the second equation (\ref{extremum}) the
"$\delta$-optimized" widths $g_\pm(\delta_{opt}(x),x)$ can be
reduced to the following rather simple form
$$
 g_\pm(\delta_{opt}(x),x)=
$$
\begin{equation}
 \label{g-opt}
 \frac{1}{2}\alpha_1''(\omega_1)-\,\widetilde{\delta}\,''_{opt}(x)
 \mp\sqrt{\big[\widetilde{\delta}\,''_{opt}(x)\big]^2+\frac{1}{4}\alpha_{12}''\,^2\,x},
\end{equation}
where
\begin{equation}
 \label{d"-opt-x}
 \widetilde{\delta}\,''_{opt}(x)=-\frac{1}{4}\Big(\alpha_2''(\omega_1)-
 \alpha_1''(\omega_1)+\alpha_2''(\omega_2)\,x\Big),
\end{equation}
 and in Eqs. (\ref{g-opt}) and (\ref{d"-opt-x}) we have put
 $\alpha_1''(\omega_2)=0$, which is true in the case
 $\omega_2<|E_1|$.

 By using Eqs. (\ref{g-opt}) and (\ref{d"-opt-x}) we can find easy
 the asymptotic expansion of the function $g_+(\delta_{opt}(x),x)$
 in powers of $1/x$ at large $x$, $x\gg 1$. The constant term in
 this expansion vanishes and the first non-zero term is given by
\begin{equation}
 \label{asympt}
 g_+(\delta_{opt}(x),x)\approx
 \frac{1}{2x}\frac{\alpha_1\,''(\omega_1)\,\alpha_2\,''(\omega_1)}
 {\alpha_2\,''(\omega_2)}.
\end{equation}

This result shows that at large $x$ the width of the narrower
quasienergy level $g_+$ decreases and tends zero as $1/x$. This is
an indication of a possibility of the unlimited narrowing of this
quasienergy level and, hence, achievement of a very high degree of
stabilization. Real limitations of narrowing and stabilization are
determined by the applicability conditions of the model. E.g., at
very large values of $x$ the second-field intensity can become too
high for the ATI processes in this field to be ignored. In
accordance with Eq. (\ref{opt-detuning}), at the optimal
conditions the increase of $x$ is accompanied by a linear increase
of the field-free detuning $\delta$. This gives other limitations
for the growth of $x$: at sufficiently large detunings the
influence of atomic levels different from $E_1$ and $E_2$ and not
taken into account in the model can become important. At last at
very large $\delta$ even the used above rotating wave
approximation can become invalid. But, on the other hand, these
limitations are not too severe, and rather significant level of
narrowing and stabilization of an atom can be reached under quite
realistic conditions. These conclusions, as well as the general
result about asymptotic decrease of the narrower-level width at
large $x$ are confirmed and specified in the following Section by
direct numerical calculations for Hydrogen and Helium atoms.

It should be noted that the curve $\delta_{opt}(x)$ includes the
point ($\delta_0,\,x_0$) where the complex detuning between the
ac-Stark shifted and broadened levels (\ref{dressed-detun}) turns
zero, $\widetilde{\delta}=0$. At this point
\begin{equation}
 \label{Gamma-tot=0}
  x_0=\frac{\alpha_1''(\omega_1)-
 \alpha_2''(\omega_1)}{\alpha_2''(\omega_2)-\alpha_1''(\omega_2)}
\end{equation}
and, in accordance with Eq. (\ref{opt-detuning}),
\begin{equation}
 \label{Delta=0}
  \delta_0=\frac{1}{4}\left\{\alpha_2'(\omega_1)-\alpha_1'(\omega_1)
 +x_0\,[\alpha_2'(\omega_2)-\alpha_1'(\omega_2)]\right\}.
\end{equation}

\section{Numerical calculations for a
$\;\;\;\;{\bf He}$ atom}

All the required numerical data are known for selected levels in
Hydrogen and Helium atoms. To a very large extent, the results
obtained for these two atoms appear to be very similar. For this
reason, below, the results of calculation of a Helium atom are
described in details, and then a short sketch of calculations for
a Hydrogen atom is given to demonstrate mainly the arising
differences and peculiarities of each atom.

\subsection{Widths of quasienergy levels}

Let us consider the following two levels of a $He$ atom and the
following two frequencies: $1s2s\equiv E_1$ and $1s4s\equiv E_2$
and $\omega_1$ = 8.44 eV (the second harmonic of a dye-laser) and
$\omega_2$ = 1,17 eV (Nd:YAG laser). For these levels and
frequencies all the polarizability tensor components (\ref{i,i}),
(\ref{off-diag}) are known \cite{Berg},\cite{Mag2}, and in atomic
units they are equal to

\noindent
\begin{equation}
 \label{data-He}
 \begin{array}{c}
 \alpha_1(\omega_1)=-\,30.42+i\,22.65,\;\;\alpha_1(\omega_2)=-236.6,\\
 \,\\
 \alpha_2(\omega_1)=-\,45.66+i\,3.21,\;\;   \alpha_2(\omega_2)=-\,479.96+i\,124.55,\\
 \,\\
 {\rm and}\;\;\alpha_{12}=38.74+i\,53.07.
\end{array}
\end{equation}

The point where the dressed-level detuning $\widetilde{\Delta}$
(\ref{dressed-detun}) (or $\widetilde{\delta}$ (\ref{relative}))
turns zero, $\widetilde{\Delta}=\widetilde{\delta}=0$, has the
following coordinates in the plane $\{x,\,\delta\}$:
\begin{equation}
 \label{Delta=0-He}
 x_0=0.156\;\;{\rm and}\;\;\delta_0=-\,13.3.
\end{equation}

The relative width of the fully dressed quasienergy levels $g_\pm$
(\ref{width-dim-less}) are shown in Fig. 6 in their dependence on
$x$ at $\delta=\delta_0$.

\vskip 12pt

\inputEps{220pt}{Fig. 6. The relative widths of quasienergy levels
of a $He$ atom $g_\pm(x)$ calculated at $\delta=\delta_0$
(\ref{Delta=0-He}).}{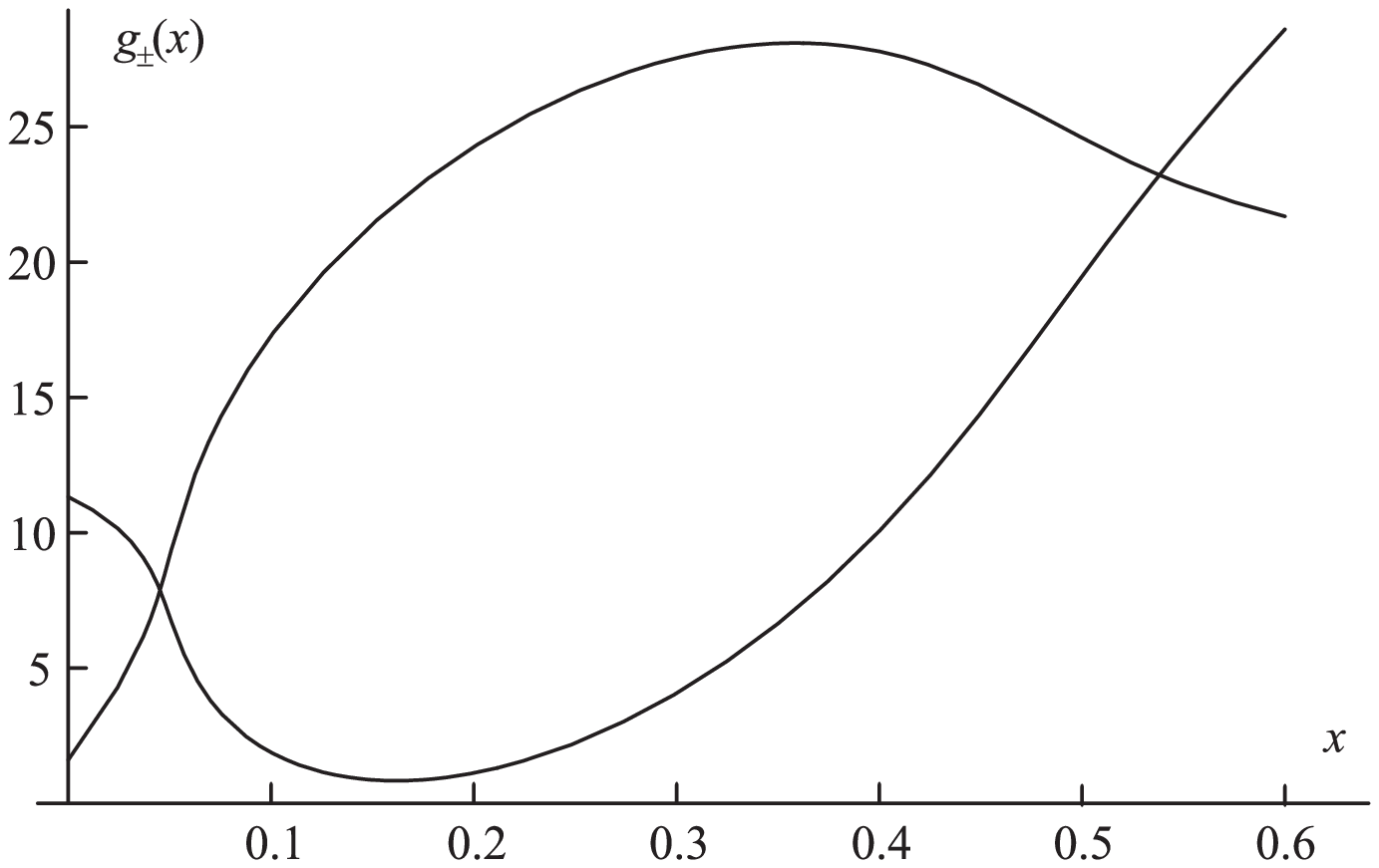}

 \noindent At this value of the
detuning $\delta$, the curves $g_+(x)$ and $g_-(x)$ cross each
other twice. The left and right crossings turn into the avoided
crossings, correspondingly, at $\delta\geq -\,11.3$ and
$\delta\leq -\,22.4$, and both crossings never turn into avoided
crossings together.

For the polarizability tensor of Eq. (\ref{data-He}), the
expression (\ref{opt-detuning}) for the optimal-narrowing detuning
$\delta_{opt}(x)$ takes the form
\begin{equation}
 \label{delta-opt-He}
 \delta_{opt}(x)=-0.26-83.57\,x.
\end{equation}

In the picture of Fig. 7 the width $g_+(\delta,\,x)$ of the
narrower quasienergy level is plotted in its dependence on the
intensity ratio $x$ at three different given values of the
field-free detuning $\delta$. The curve at $\delta=-11$ differs

\vskip 21pt

\inputEps{220pt}{Fig. 7. Three typical curves $g_+(\delta=const,\,x)$ for
$He$ calculated at $\delta=-11,\,-15,\,{\rm and}\,-23$ (from the
left to the right)}{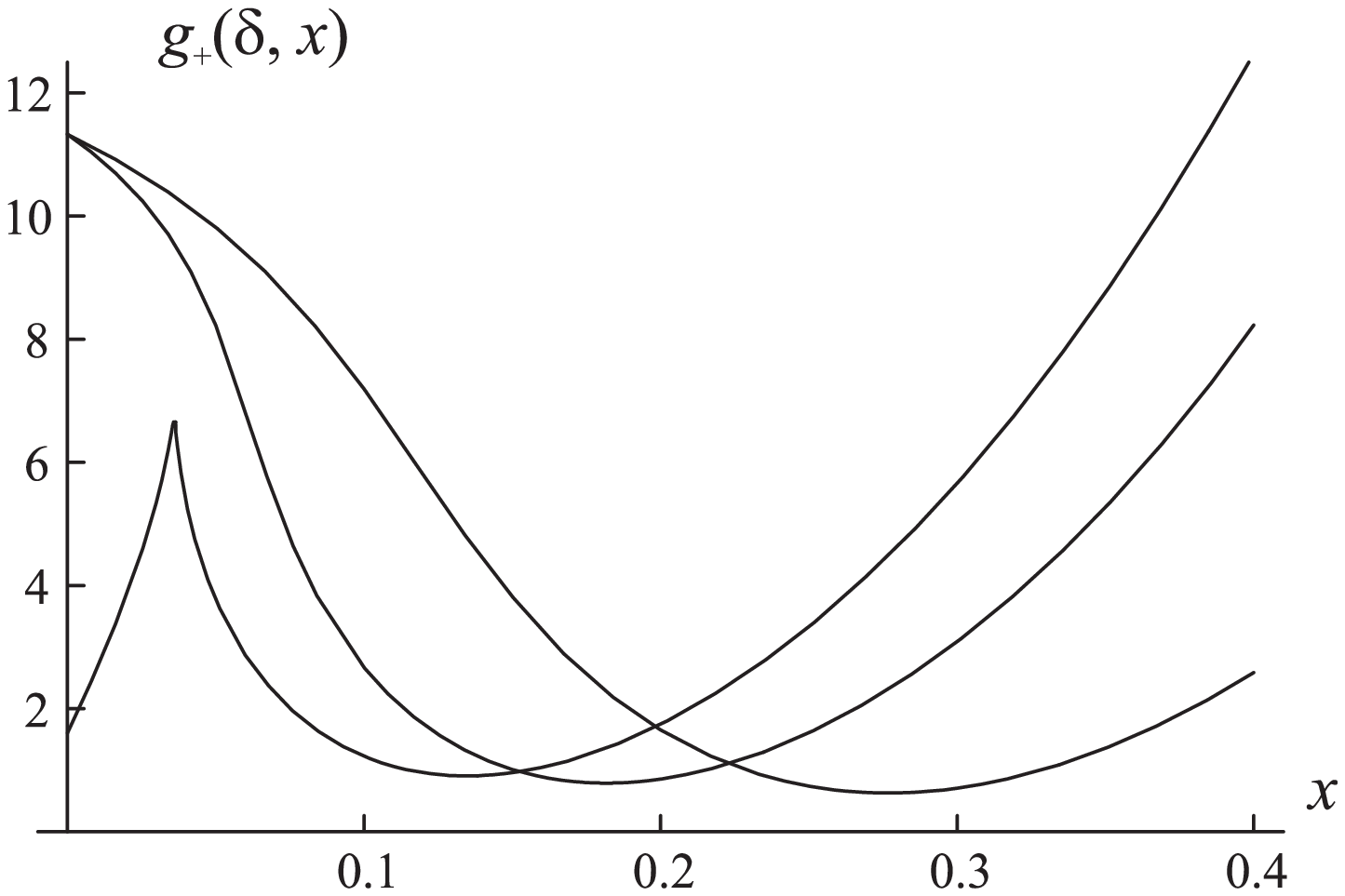}

\noindent qualitatively from two other curves. The difference
arises because for this curve the detuning $\delta$ is large
enough for the left crossing of Fig. 6 to turn into the avoided
crossing. The sharp peak of the curve  $g_+(\delta=-11,\,x)$ at
Fig. 7 is an indication of the root-square branch-point-like
behavior, which takes place at $\delta=-11.3$ and which is similar
to the root-square branch-point behavior of the curve of Fig. 1
for $\gamma_\pm''$ in an idealized two-level system of Section
${\bf 1A}$.

The minima of all three curves at Fig. 7 are seen to be getting
the deeper the larger is $|\delta|$. Positions of these minima,
$x^{min}(\delta)$, are determined by the optimal detuning
(\ref{delta-opt-He}) and can be found from the equation
$\delta=\delta_{opt}(x)$. The $x-$dependent width of the narrower
quasienergy level, minimized with respect to $\delta$, is given by
$g_+^{min}(x)\equiv g_+(\delta_{opt}(x),\,x)$, and its dependence
on $x$ is given by the curve of Fig. 8.

\vskip 12pt

\inputEps{220pt}{Fig. 8. The width of the narrower quasienergy level
 of a Helium atom $g_+^{min}(x)$, minimized with respect to the
detuning $\delta$, in its dependence on $x$.}{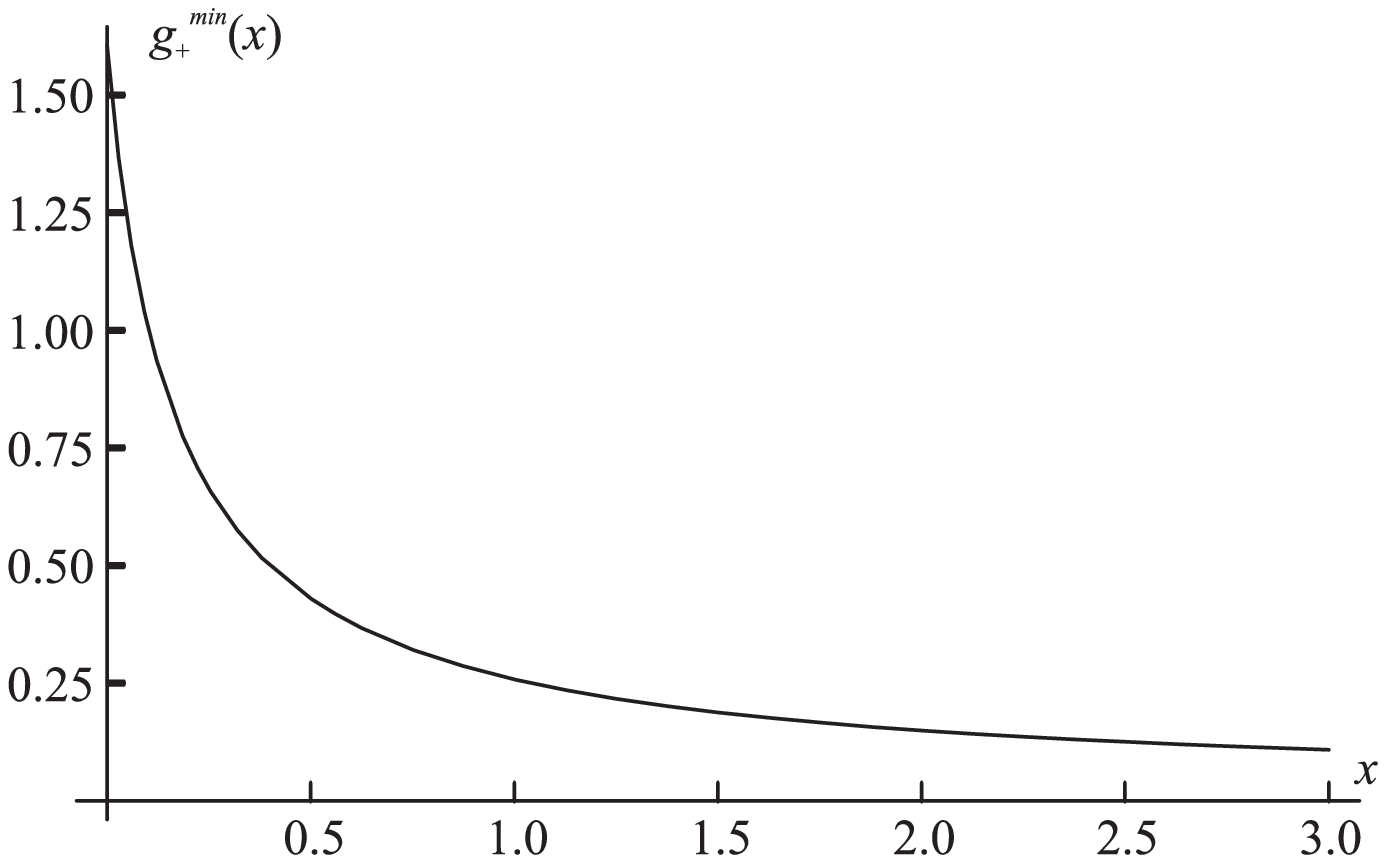}

A monotonous fall of the function $g_+^{min}(x)$ means that the
$x-$dependent width of the narrower quasienergy level, minimized
with respect to the detuning $\delta$, tends to zero at
asymptotically large values of $x$. In other words, in the
framework of the used model one of the two quasienergy levels of
the system can be narrowed unlimitedly by means of increasing the
ratio of intensities $x=I_2/I_1$ and the field-free detuning
$|\delta|$ in such a way that the equation
$\delta=\delta_{opt}(x)$ remains satisfied. Real limitations of
such a narrowing are determined only by the model applicability
conditions: (i) at very large values of $x$ the second-field
intensity $I_2$ will become too high to ignore above-threshold
ionization produced by this field and (ii) at very large values of
$|\delta|$ other levels ignored above can become more important
and, at $|\Delta|\sim \omega_2$, the rotating-wave approximation
can become invalid. But at sufficiently long pulse durations
$\tau$ and low first-field intensity $I_1$ these limitations are
not too severe, and the achievable degree of narrowing can be
rather high.

\subsection{Probabilities of ionization and non-ionization}

The residual probabilities to find a $He$ atom in its bound states
after interaction with a two-color field in the case of a
rectangular envelope is determined by Eqs. (\ref{w-res}),
(\ref{w-1-expl}), (\ref{w-2-expl}), and the results of the
corresponding calculations are shown in Fig. 9. The three
resonance-like

\vskip 12pt

\inputEps{220pt}{Fig. 9. The residual probability to find a $He$ atom
in bound states calculated at $\delta= -100,\,-250,\,{\rm
and}\,-500$ (from the left to the right) and
$\delta=\delta_{opt}(x)$ (\ref{delta-opt-He}) (the upper curve),
$\theta=0.1$. }{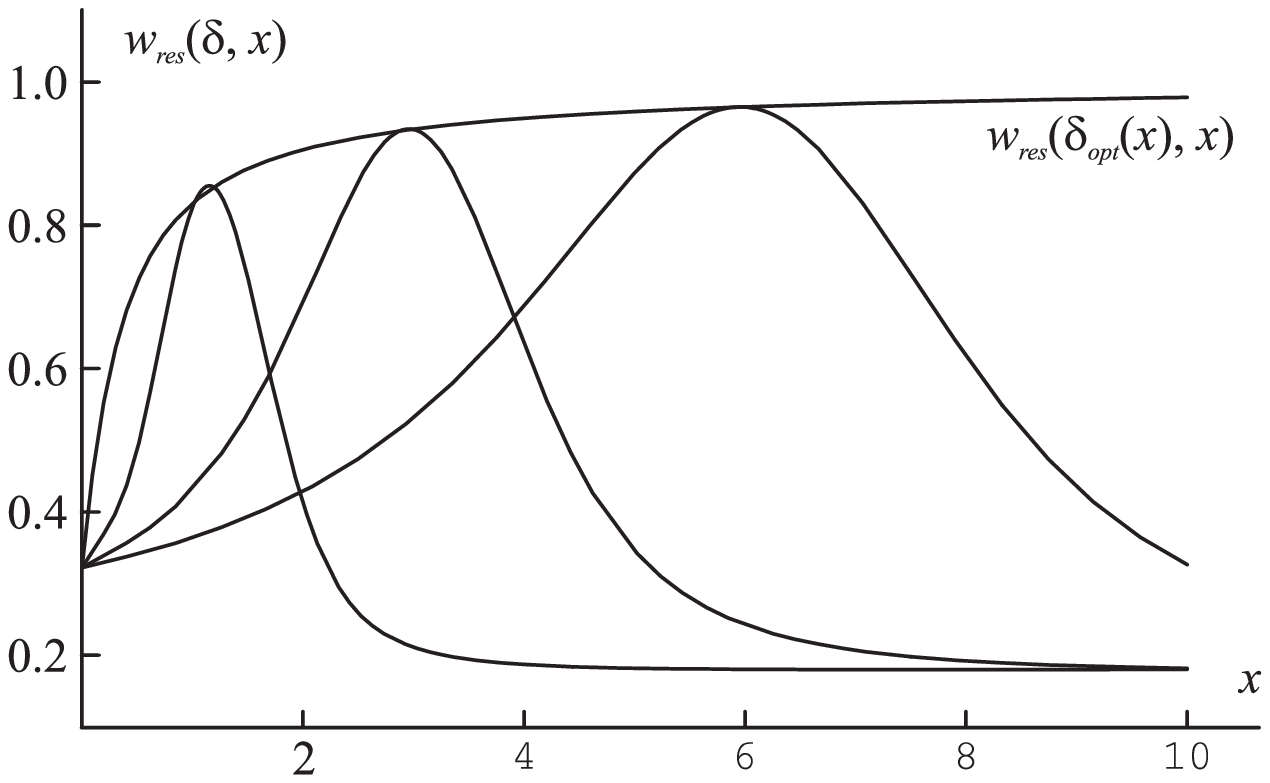}

\noindent curves correspond to three different values of the
field-free detuning $\delta$. The envelope of peaks of these
curves is the maximized residual probability equal to
$w_{res}(\delta_{opt}(x), x)$ with $\delta_{opt}(x)$ given by Eq.
(\ref{delta-opt-He}). These results show that the function
$w_{res}(\delta_{opt}(x),x)$ monotonously grows approaching one at
large values of $x$. This means  that under optimal conditions
stabilization of a $He$ atom in a two-color field can be very
high, more than 90$\%$.

The picture of Fig. 10 shows the distribution of the residual
probability between the levels $E_1$ and $E_2$.

\vskip 12pt

\inputEps{220pt}{Fig. 10. The ratio $w_2/w_1$ for a $He$ atom
at $\delta=\delta_{opt}(x)$ (\ref{delta-opt-He}),
$\theta=0.1$.}{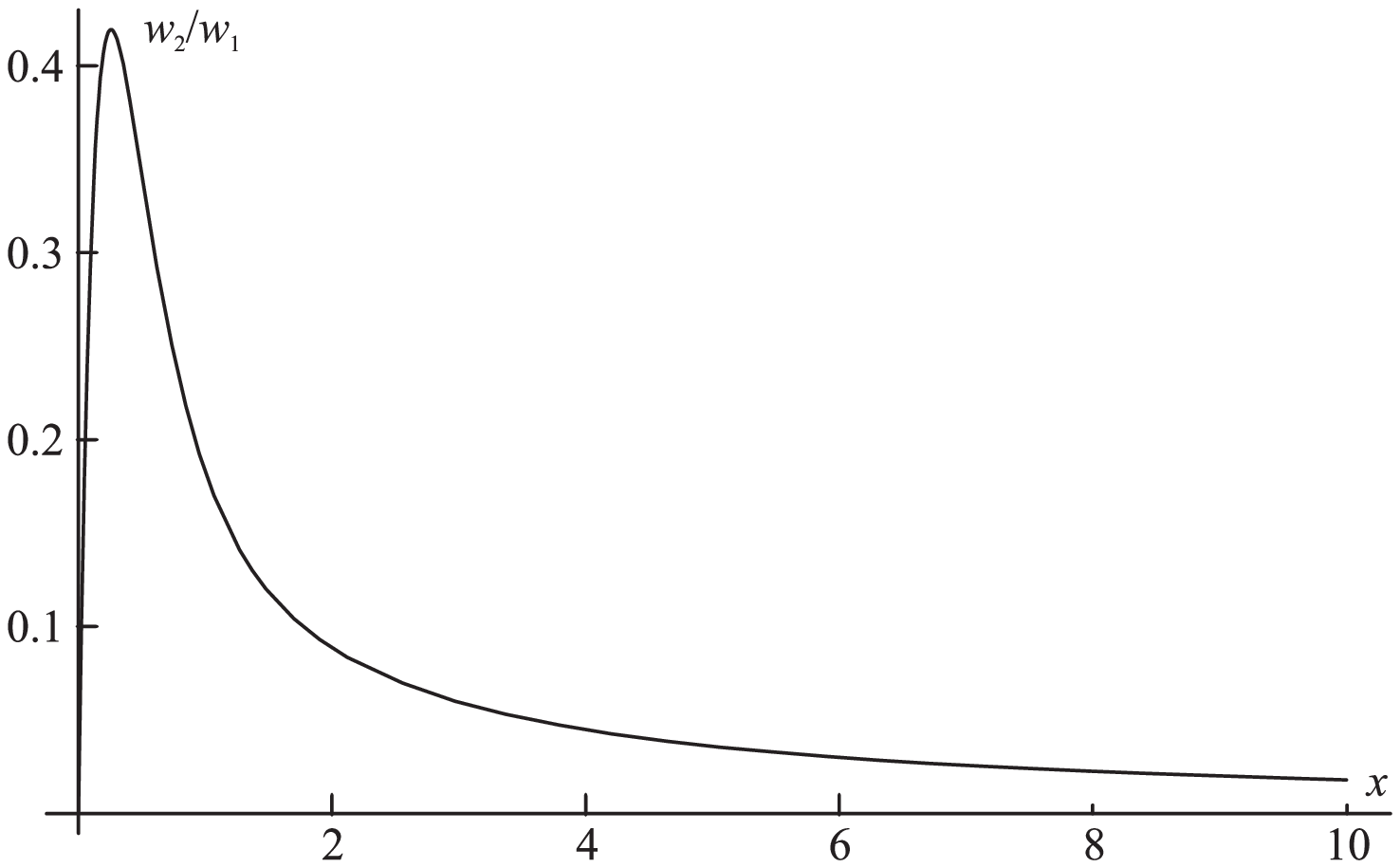}

\noindent Under the conditions of optimal stabilization
($\delta=\delta_{opt}(x)$) at sufficiently high value of the
intensity ratio $x$, the ratio of probabilities $w_1/w_2$ falls
tending asymptotically to zero. This means that under the optimal
stabilization conditions, interference suppresses not only
ionization but also excitation of the level $E_2$, which can be
seen experimentally also.

In all pictures of this Section (Figs. 6-10) the calculated values
are plotted in their dependence on the intensity ratio $x$. To see
such dependencies in experiments one has to make, for example, a
series of measurements at different values of the second-field
intensity $I_2$ at a given first-field intensity $I_1$. Another
possible way of an experimental investigation is keeping the ratio
$x=I_2/I_1$ constant and changing both intensities synchronously.
Calculated for such a scheme of measurements, the probability of
ionization in its dependence on $I_1\propto I_2$ is given by the
curves of Fig. 11. In this picture the intensity $I_1$ is
expressed in units of an arbitrary constant intensity $I_0$ at
$x=I_2/I_1=3$. The detuning $\Delta$ and pulse duration $\tau$ are
taken to be equal to $\Delta=-200\times I_0$ and $\tau=0.1/I_0$
($a$) or $\tau=1/I_0$ ($b$), where $\Delta$, $\tau$, and $I_0$ are
in atomic units.

Normalization by an arbitrary constant $I_0$ reflects the scaling
effect described above in Section ${\bf V}$. A possibility to
choose any value of $I_0$ indicates a large flexibility of the
system under consideration with respect to a choice of of the
light intensities and pulse durations. For example, if
$I_0=10^{-6}$a.u. ($\approx 3\times 10^{10}$ W/cm$^2$), the
intensity $I_2=3\times I_1$ does not exceed $3\times 10^{11}$
W/cm$^2$ in the variation range of $I_1$ at Fig. 11, which is low
enough for no ATI effects to take place. And, under the best
stabilization conditions, the detuning $\Delta$ and pulse duration
are equal to $\Delta=-240\times I_0=-2.4\times 10^{-4}\,{\rm
a.u.}\sim 0.065\, {\rm eV}\ll\omega_{1,\,2}$ and
$\tau=1/I_0=10^6{\rm a.u.}\sim 3\,{\rm ps}$.

\vskip 12pt

\inputEps{180pt}
{Fig. 11. $w_i(I_1/I_0)$ at $I_2=3I_1$, $\delta=-300\times
(I_0/I_1)$ and $\theta=0.1\times(I_1/I_0)\,(a)$ and
$I_1/I_0\,(b)$.}{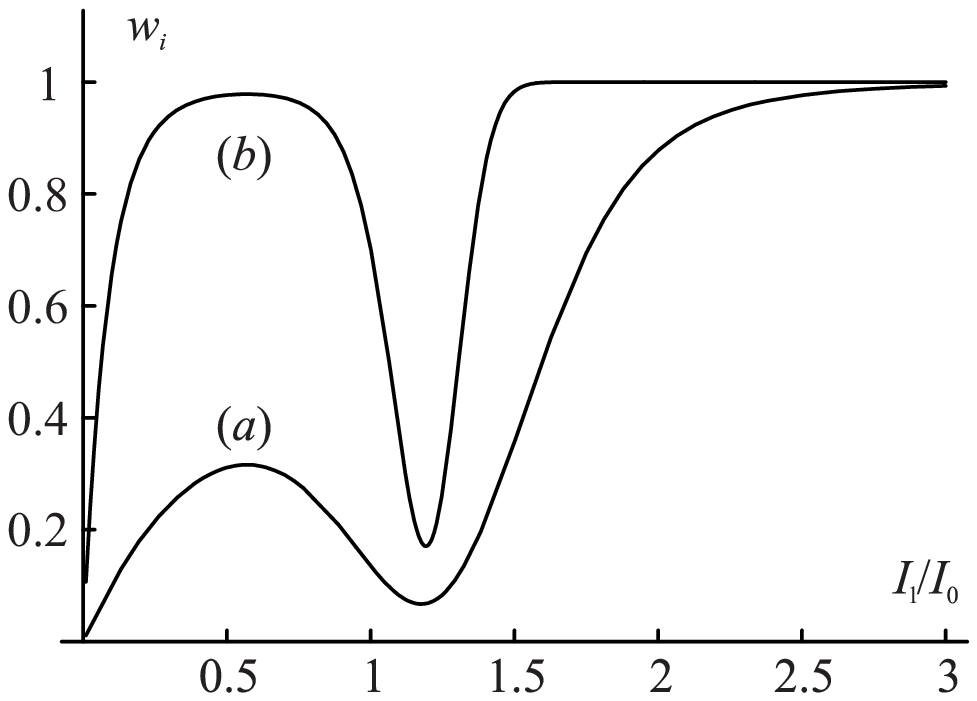}

The curves of Fig. 11 are typical for the stabilization picture.
With a growing light intensity, at first, the probability of
ionization grows (perturbation theory region), then falls, and
this is the beginning of the stabilization window, and finally
grows again, which corresponds to the break of stabilization.
Stabilization and its break arise because, owing to the ac Stark
shift and level mixing, with a growing light intensity the system
comes to and, then, goes out of the resonance conditions, optimal
for fully-dressed-level narrowing and interference stabilization.
The curve $b$ of Fig. 11 indicates an appearance of an additional
region between the perturbation theory and stabilization zone at a
sufficiently long pulse duration $\tau$. This is the region of the
total ionization of an atom, where $w_i=1$. This means that at
some intermediate intensity light pulses provide a complete
ionization and at a stronger field, owing to interference,
ionization becomes rather small ($w_i\sim 0.2$). In absolute
values, the minimal achievable probability in the stabilization
region is somewhat lower in the case of short pulses
($\tau=0.1I_1/I_0$) (the curve $a$ of Fig. 11). But the degree of
stabilization can be determined alternatively as the ratio of the
maximal probability achievable in the region between
perturbation-theory and stabilization regions to the minimal value
of $w_i$ in the stabilization window. In terms of such a
definition, the degree of stabilization is much higher in the case
of longer pulses ($\tau=I_1/I_0$) (the curve $b$ of Fig. 11). Of
course, a further increase of the pulse duration flattens the
curve of Fig. 11 in the stabilization region and decreases the
degree of stabilization. In this sense, the pulse duration chosen
for the curve $b$ of Fig. 11, $\tau=I_1/I_0$,  is close to the
optimal one.

The picture of Fig. 12 characterizes spectral features of the
residual probability to find an atom in its bound states at the
best stabilization conditions of Fig. 11: $I_2/I_1=3$,
$I_1/I_0=1.2$, $\theta=0.12$ ($a$) and $1.2$ ($b$) (which
corresponds to $\tau=0.1/I_0$ and $\tau=1/I_0$).

\vskip 12pt

\inputEps{180pt}{Fig. 12. $w_{res}(\delta)$ at $I_2=3I_1$,
$I_1=1.2I_0$, $\theta=0.12\,(a)$ and $1.2\,(b)$.}{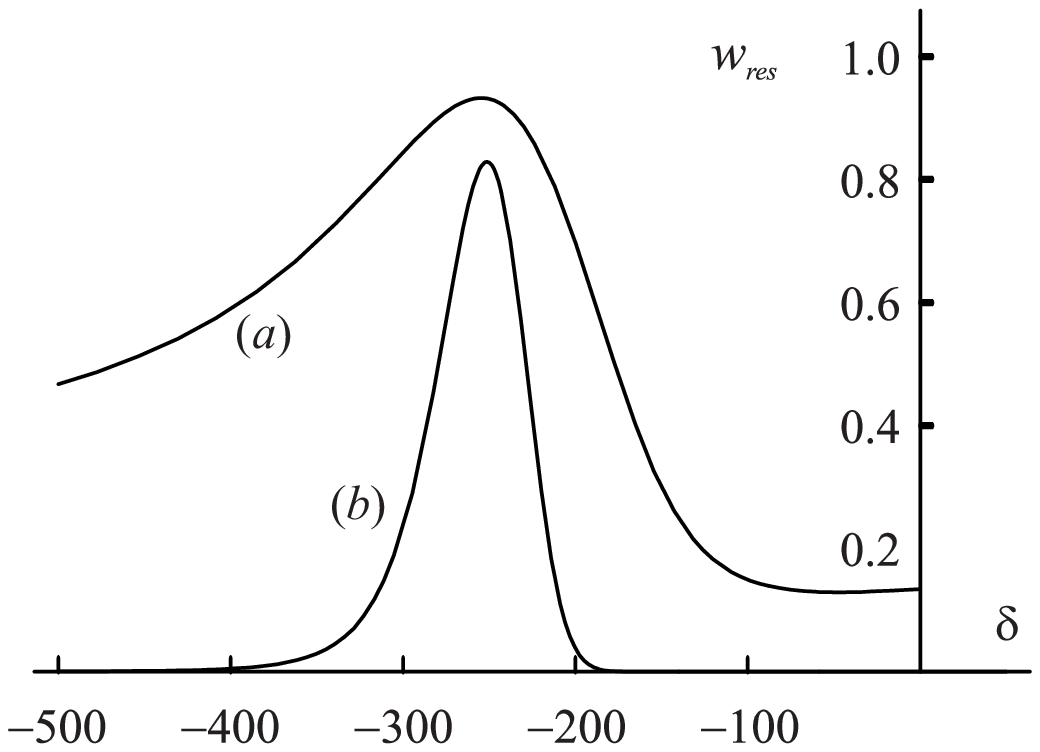}

The curve $a$ of this picture looks similar to the Fano curve of
Fig. 1. This shows that at $\tau=0.12/I_0$, still, the effect of
stabilization under discussion can be interpreted as a
strong-field LICS. But in the case of longer pulses,
$\tau=1.2/I_0$ (the curve $b$ of Fig. 12) similarity with LICS
practically disappears. The only reminder about a remote
connection with the Fano curve is a slight asymmetry of the curve
$b$ of Fig. 12. Apart from this, the curve $b$ describes the
effect of interference stabilization in its pure form.

\subsection{Smooth envelope}

All the results described above were derived for pulses with a
rectangular envelope. Usually, envelopes of short laser pulses are
smooth. To consider such a more realistic situation, we have
solved general equations (\ref{2-lev-eq-gen}) with the $\sin^2$
pulse envelopes (\ref{pure-sin^2}). The results of such a solution
are shown in Fig. 13, which is a direct analog of Fig. 9. Again, a
series of resonance-like curves describes the residual probability
of finding an atom in its bound states $w_{res}(\delta, x)$ at
various given values of the detuning $\delta$  and the intensity
ratio $x$ considered as the independent variable. The residual
probability maximized with respect to the detuning $\delta$ is the
envelope of the peaks of these curves. In Fig. 13 such a maximized
probability is approximated by the functions
$w_{res}(\delta_{opt}^{\sin}(x), x)$, where
$\delta_{opt}^{\sin}(x)$ is the empirically found linear function
providing the best fitting to the peak envelope:
\begin{equation}
 \label{delta-opt-smooth-He}
 \delta_{opt}^{\sin}(x)=3.122-73.3\, x
\end{equation}

\vskip 12pt

\inputEps{220pt}{Fig. 13. A series of curves $w_{res}(\delta, x)$
at various given values of the detuning $\delta$ and the function
$w_{res}(\delta_{opt}^{sin}(x), x)$ (a thick curve) for a $He$
atom and the $\sin^2$ pulse envelopes (Eq.
(\ref{pure-sin^2}).}{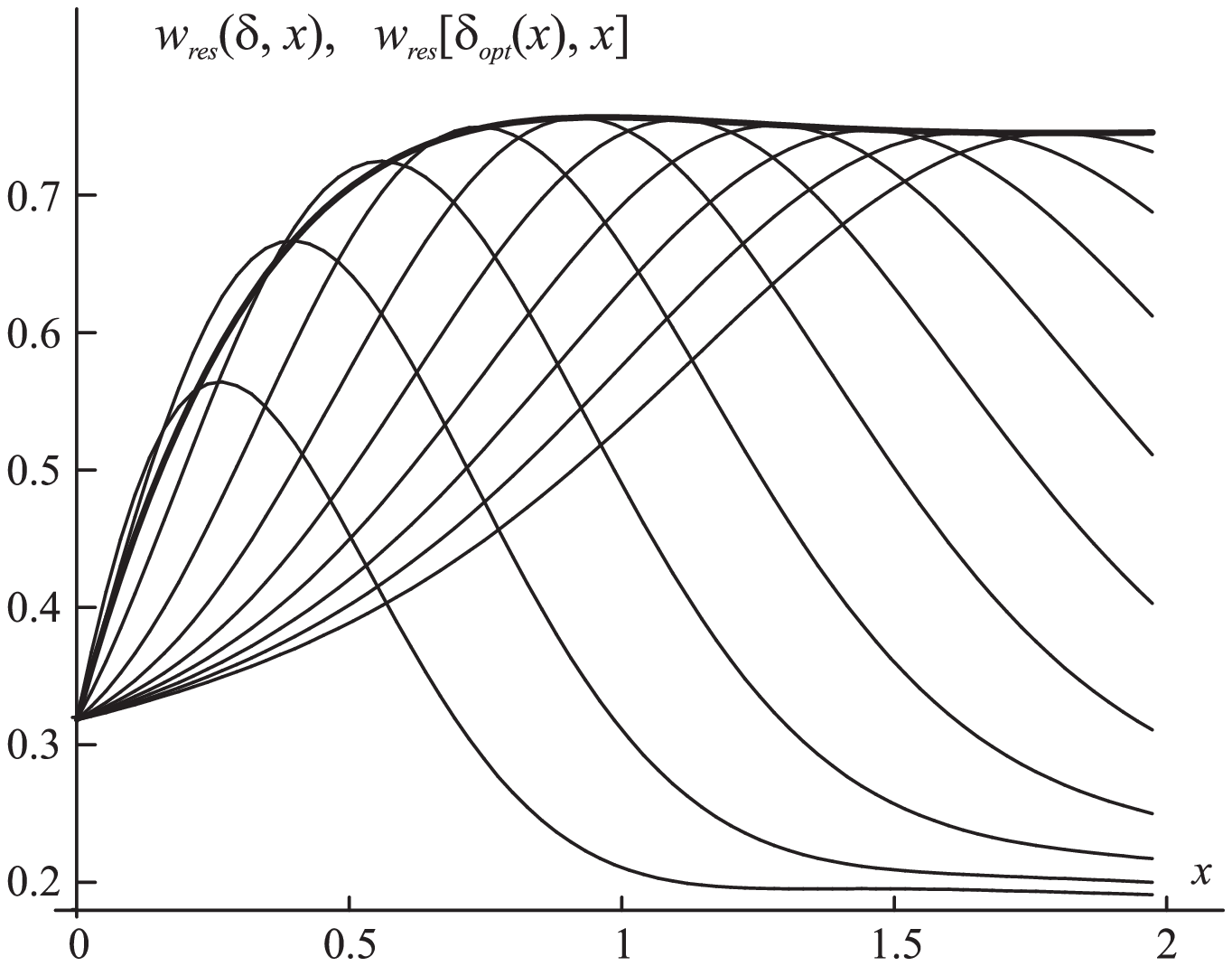}

In Fig. 14 we plot the maximized residual probability to find $He$
atoms in bound states calculated in the cases of rectangular (the
curves 1 and 2) and $\sin^2$ (the curve

\vskip 12pt

\inputEps{180pt}{Fig. 14. The functions
$w_{res}^{rect}(\delta_{opt}^{rect}(x),x)$ (1),
$w_{res}^{rect}(\delta_{opt}^{\sin}(x),x)$ (2), and
$w_{res}^{\sin}(\delta_{opt}^{\sin}(x),x)$ (3) with
$\delta_{opt}^{rect}(x)$ and $\delta_{opt}^{\sin}(x),x$ given by
Eqs. (\ref{delta-opt-He}) and (\ref{delta-opt-smooth-He}),
respectively.}{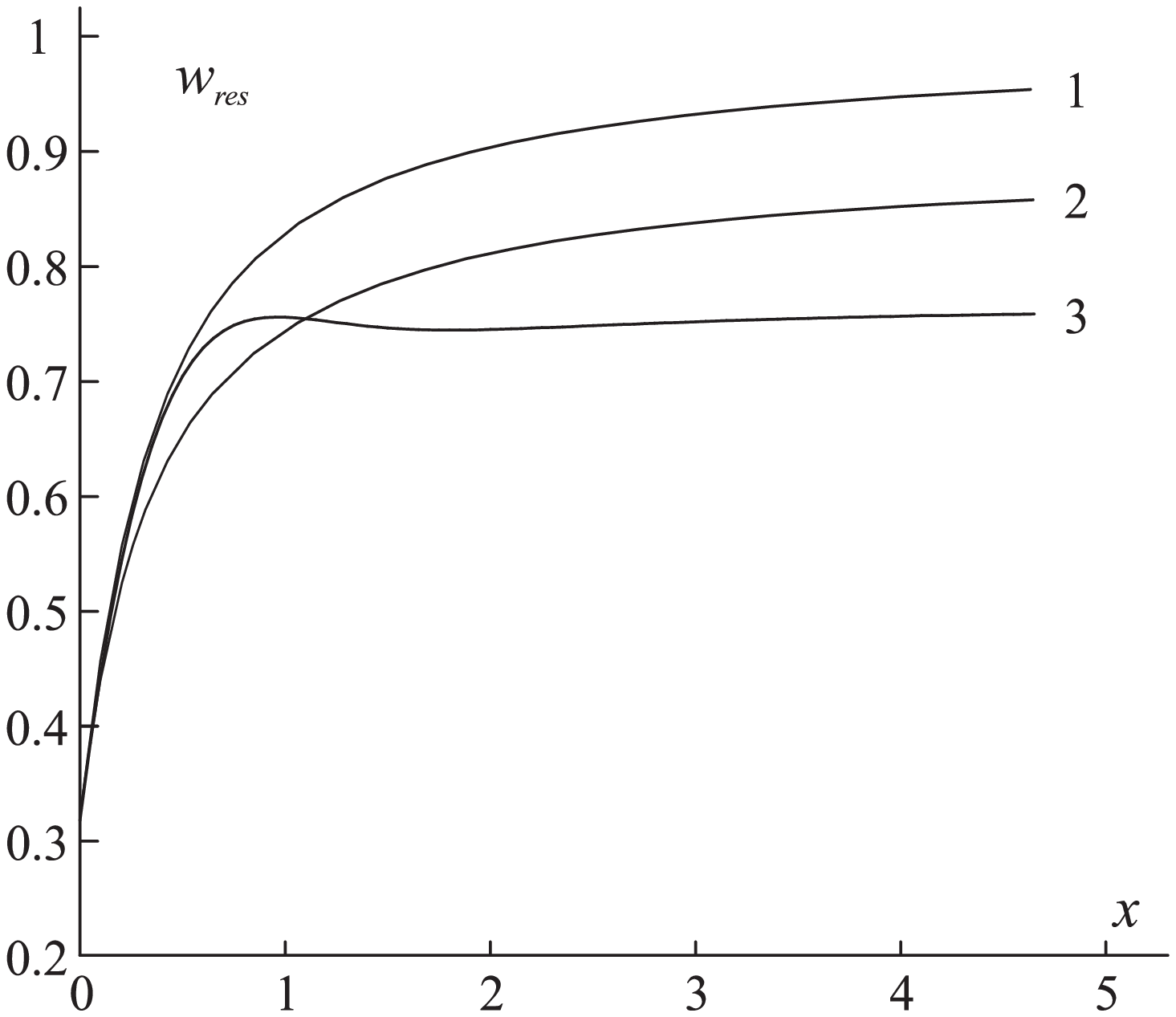}

\noindent 3) pulse envelopes at $\delta=\delta_{opt}^{rect}(x)$
(\ref{delta-opt-He}) (the curve 1) and
$\delta=\delta_{opt}^{\sin}(x)$ (\ref{delta-opt-smooth-He}) (the
curves 2 and 3). Comparison of the curve 1 and 3 shows that a
transition to a smooth envelope $\varepsilon_0(t)$ reduces a
little bit the maximal achievable degree of stabilization compared
to the rectangular-envelope case, but not too much (70-75 $\%$
instead of $90\%$). Moreover, the curve 3 of Fig. 14 shows that in
the case of a smooth pulse envelope the residual probability
remains more or less stable in a rather large variation interval
of the  intensity ratio $x$, approximately from 0.5 to 5 and more.
This shows that the effect of stabilization is rather robust.

Another interesting effect seen rather well from  comparison of
the curves 2 and 3 of Fig. 14. These two curves are calculated at
the coinciding dependencies of the detuning $\delta$ on the
intensity ratio parameter $x$, $\delta=\delta_{opt}^{\sin}(x)$
(\ref{delta-opt-smooth-He}). As it's seen well from Fig. 14, at
$x\leq 1$ the curve 3 goes above the curve 2. This means that at
the same detunings the residual probability to find an atom in its
bound states in the case of smooth envelope pulses exceeds the
same probability at a rectangular pulse envelope. In other words,
in this range of the intensity ratio parameter $x$ smoothing of
the pulse envelope increases rather than reduces the degree of
stabilization. This conclusion follows directly from calculations
though it looks counterintuitive and, in this sense, rather
interesting.

\section{Hydrogen}

In a hydrogen atom, all the polarizability tensor components are
known for the levels $2s$ and $5s$ and frequencies $\omega_1$ =
4.02 eV ($XeCl$ laser) and $\omega_2$ = 1.17 eV ($Nd:YAG$ laser)
\cite{Mag}. In atomic units they are given by
\begin{equation}
 \label{data-H}
 \begin{array}{c}
 \alpha_1(\omega_1)=-\,45.56+i\,27.29,\;\;\alpha_1(\omega_2)=179.92,\\
 \,\\
 \alpha_2(\omega_1)=-\,45.66+i\,1.78,\;\;\alpha_2(\omega_2)=-\,513.76+i\,93.83\\
 \,\\
 {\rm and}\;\;\alpha_{12}=6.56+i\,50.60.
\end{array}
\end{equation}
The data (\ref{data-H}) show, in particular, that
$\Gamma_2^{(1)}/\Gamma_1
=\alpha_2^{\prime\prime}(\omega_1)/\alpha_1^{\prime\prime}(\omega_1)\sim
6.5\times 10^{-2}$, which means that the assumption
(\ref{smaller}) is pretty well satisfied.

Rigorously, in a hydrogen, there are other levels ($5d$ and $5f$)
of almost the same energy as $5s$. Owing to the selection rules,
the level $5f$ is not connected either with $2s$ or $5s$ levels by
two-photon Raman-type transitions and, for this reason, can be
ignored. As for the level $5d$, in principle, it can participate
in a scheme of two-photon Raman-type transitions under
consideration. However, by ignoring at first this additional level
let us consider a two-level 2s-5s scheme, analogous to that of the
previous Section.

For this system the  coordinates of the $\widetilde{\delta}=0$
point in the $\{x,\,\delta\}$ plane are given by
\begin{equation}
 \label{H-opt}
 x_0=0.272\;\;{\rm and}\;\;\delta_0=-\,47.2.
\end{equation}
At $\delta=\delta_0$, the calculated relative width of the fully
dressed quasienergy levels $g_\pm$ (\ref{width-dim-less}) in their
dependence on $x$ are shown in Fig. 15.

Compared with Fig. 5, the picture of Fig. 15 indicates the first
well pronounced difference between Helium and Hydrogen. In the
case of a Hydrogen the curves of widths of quasienergy levels vs.
$x$ have two avoided-crossing points whereas in the case of Helium
such a situation never occurs and at $\delta=\delta_0$ there are
two real-crossing points (Fig. 5).

\vskip 12pt

\inputEps{180pt}{Fig. 15. The functions $g_\pm(\delta_0,x)$
(\ref{relative})} {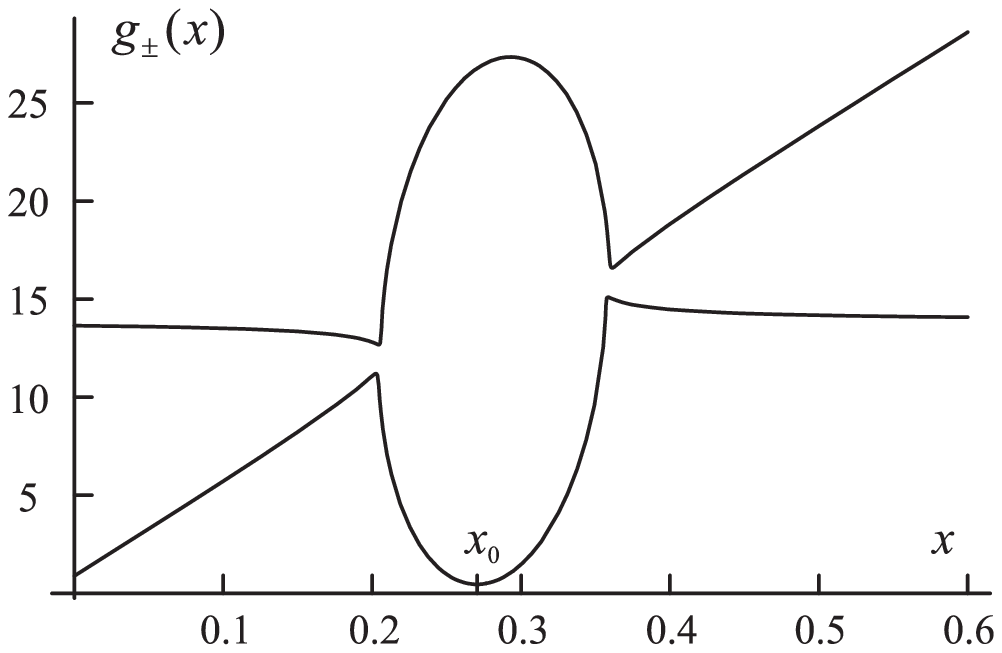}

Another important difference concerns smooth pulse envelopes and
the third level effect. To solve such a problem, we have to
generalize Eqs. (\ref{exp2}) and (\ref{2-lev-eq-gen}). In Eq.
(\ref{exp2}), in accordance with the more general Eq.
(\ref{expansion}), there appears an additional term
$C_3(t)\,e^{i\omega_2t}\,\psi_3$. Then, equations for $C_i(t)$
($i=1,\,2,\,3$) take the form
\begin{equation}
 \label{3-lev-eq-gen}
 \begin{array}{c}
 i{\dot C}_1-(\widetilde{E}_1(t)+\omega_1)\,C_1\\
 =-\frac{1}{4}\varepsilon_{1\,0}(t)\varepsilon_{2\,0}(t)
 \Big(\alpha_{12}C_2+\alpha_{13}\,C_3\Big),\\
 \,\\
 i{\dot C}_2-(\widetilde{E}_2(t)+\omega_2)\,C_2
 =-\frac{1}{4}\varepsilon_{1\,0}(t)\varepsilon_{2\,0}(t)\alpha_{21}\,C_1\\
 -\frac{1}{4}\Big(\alpha_{23}(\omega_1)\varepsilon_{1\,0}^2(t)+
 \alpha_{23}(\omega_2)\varepsilon_{2\,0}^2(t)\Big)\,C_3(t),\\
 \,\\
 i{\dot C}_3-(\widetilde{E}_3(t)+\omega_2)\,C_3=
 -\frac{1}{4}\varepsilon_{1\,0}(t)\varepsilon_{2\,0}(t)\alpha_{13}\,C_1\\
 -\frac{1}{4}\Big(\alpha_{23}(\omega_1)\varepsilon_{1\,0}^2(t)+
 \alpha_{23}(\omega_2)\varepsilon_{2\,0}^2(t)\Big)\,C_2(t)
 \end{array}
\end{equation}
where $\widetilde{E}_3$ is given by the same Eq.
(\ref{dressed-levels}) as $\widetilde{E}_1$ and $\widetilde{E}_2$
with $\alpha_3(\omega_{1,\,2})$ and new off-diagonal elements of
the polarizability tensor given by \cite{Mag}
\begin{equation}
 \label{new-alpha}
 \begin{array}{c}
 \alpha_3(\omega_1) = -43.26 + i\,0.42,\;\; \alpha_3(\omega_2) =
 -405.9+i\,81.8,\\
 \,\\
 \alpha_{23}(\omega_1) = 0.74+i\,0.21,\;\;\alpha_{23}(w_2)=68.61+i\,21.63,\\
 \,\\
 {\rm and}\;\;\;\alpha_{13} = 6.15+i\,11.69.
 \end{array}
\end{equation}

Found from Eqs. (\ref{3-lev-eq-gen}) and (\ref{sin^2}) the
residual probability to find an atom in its bound states is shown
in Fig. 16 for three different values of the pulse envelope
smoothing factor $a$.

Two rather interesting conclusions can be deduced from this
picture. First, a strong smoothing of pulse envelopes decreases
the peak value of the residual probability to find an atom in
bound states. At the chosen value of the detuning $\delta=-530$ in
the case of pure $\sin^2$ pulses $[w_{res}(x)]_{max}$ is almost
twice smaller then in the case of rectangular envelopes. The
smoothing induced decrease of the residual probability in the case
of Hydrogen is much more significant than in the case of Helium.

\vskip 35pt

\inputEps{220pt}{Fig. 16. The function $w_{res}(x)$ in a three-level
scheme at $\delta=-530$, $\theta=0.1$, and the envelope smoothing
parameter in (\ref{sin^2}) $a=100,\,10,\,{\rm and}\,0.1$ (from top
to bottom) .}{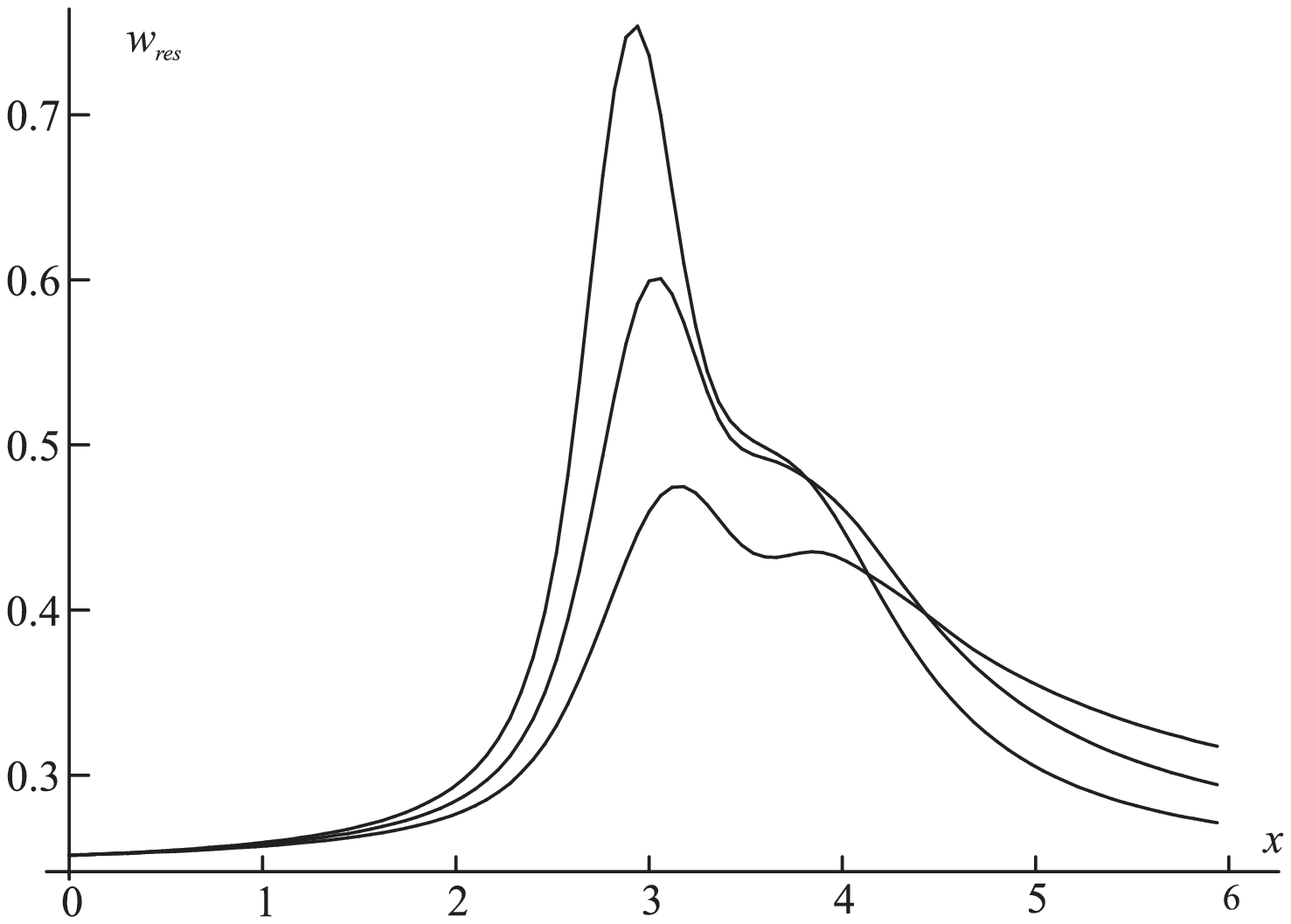}

The second effect seen in the picture of Fig. 16 concerns the
influence of the third level. This influence manifests itself in a
shoulder on the curves $w_{res}(x)$, but the third level is seen
not to affect much the main maximum of the curves.

$$\,$$
\section{Conclusion}

To summarize, we describe and discuss a scheme of interaction of
atoms with radiation of two lasers. Intensity and pulse duration
of lasers are assumed to be high and long enough to provide full
ionization in the field of each of these two lasers alone if only
atoms are prepared initially at levels from which one-photon
ionization can take place. We show that owing to interference
effects under the conditions close to Raman-type resonance between
some two selected atomic levels ionization of an atom experiencing
a joint action of the field of two lasers can be significantly (up
to 90 $\%$) suppressed. Optimization of such a stabilization
effect involves optimization with respect to the Raman-type
resonance detuning and the ratio of the two laser intensities.
Specific calculations are carried out for Hydrogen and Helium
atoms for couples of atomic levels and laser frequencies at which
information about the complex polarizability tensors involved is
available. Qualitatively, the results of calculations for Hydrogen
and Helium appear to be very similar. This gives us a reason to
think that the effect described is rather universal, and can occur
also at other atoms, levels, and frequencies. The dependence of
the effect on laser pulse shapes is investigated. It is shown that
in the case of Helium atoms sensitivity of the results to a pulse
shape is lower than in the case of Hydrogen atoms. In Helium, even
in the case of smooth pulses, the degree of stabilization remains
rather high (more than 70 $\%$), and the effect exists at this
level in a rather large range of the intensity ratio parameter
$x$. The described scaling effect gives a possibility to select
ranges of variation of the laser pulse peak intensities and pulse
duration in a ranges most convenient for experimental observation.

\section*{ACKNOWLEDGMENT}

The work is supported partially by RFBR grants 02-02-16400 and
03-02-06144.


\begin{thebibliography}{99}

\bibitem{Mov} M.V. Fedorov and A.M. Movsesian {\em J. Phys. B}, {\bf 21},
L155 (1988)
\bibitem{Book} M.V. Fedorov {\em Atomic and Free Electrons in a Strong
Light Field}, World Scientific: Singapore, 1997.
\bibitem{Armstr} L. Armstrong, Jr., B. Beers, and S. Feneuille {\em Phys. Rev. A}, 12, 1903
(1975)
\bibitem{Heller}  Yu.I. Heller and A.K. Popov {\em Opt. Commun.}, {\bf 18}, 449
(1976)
\bibitem{Andr} A.I. Andryushin and M.V. Fedorov {\em Izvestiya Vuzov:
Fizika}, $\# 1$, 63 (1978)
\bibitem{Lambr} Bo-nian Dai and P. Lambropulos {\em Phys. Rev. Lett.}, {\bf 36}, 5202
(1987)
\bibitem{PQE} M.V.Fedorov and A.E.Kazakov {\em Progress in Quantum Electronics},
{\bf 13}, 97 (1989)
\bibitem{Knight} P.L. Knight et al. {\em Phys. Rep.}, {\bf 190}, 1 (1990)
\bibitem{Mag} A.I. Magunov, I. Rotter, and S.I. Strakhova {\em J. Phys. B},
{\bf 34}, 29 (2001)
\bibitem{Hutch} M.H.R. Hutchinson and K.M.M. Ness {\em Phys. Rev. Lett.}, {\bf 60}, 105 (1988)
\bibitem{Shao} Y.L. Shao et al. {\em Phys. Rev. Lett.}, {\bf 67}, 3669 (1991)
\bibitem{Caval} S. Cavalieri and F. S. Pavone {\em Phys. Rev. Lett.}, {\bf 67}, 3673 (1991)
\bibitem{Half} T. Halfmann et al. {\em Phys. Rev. A}, {\bf 58}, 46 (1998)
\bibitem{Eberly} K. Rzazewski and J.H. Eberly {\em Phys. Rev. Lett.
A}, {\bf 47}, 408 (1981)
\bibitem{Lambr2} P. Lambropoulos and P. Zoller {\em Phys. Rev.
A}, {\bf 24}, 379 (1981)
\bibitem{AKF} A.I. Andryushin, M.V. Fedorov, and A.E. Kazakov {\em
J. Phys. B}, {\bf 15}, 2851 (1982)
\bibitem{Rahman} A. Lami and N.K. Rahman {\em Phys. Rev. A}, {\bf
26}, 3360 (1982); {\bf 33}, 782 (1986); {\bf 34}, 3908 (1986);
{\bf 40}, 2385 (1989)
\bibitem{Gozzini} G. Alzetta, A. Gozzinin, L. Moi, and G. Orriols
{\em Nuovo Cimento B}, {\bf 36}, 5 (1976)
\bibitem{Ari-Orri} E. Arimondo E. Arimondo and G. Orriols {\em
Nuovo Cimento Lett.}, {\bf 17}, 333 (1976)
\bibitem{Stroud} R.M. Whitley and C.R. Stroud {\em Phys. Rev. A},
{\bf 14}, 1498 (1976)
\bibitem{Arimondo} E. Arimondo {\em Progress in Optics}, Ed. E. Wolf, {\bf 35}, 257 (1996)
\bibitem{Berg} L.P. Yatsenko, T. Halfman, B.W. Shore, and K. Bergmann, {\em Phys.
Rev. A}, {\bf 59} (1999)
\bibitem{Mag2} A.I. Magunov and S.I. Strakhova, {\em Quantum
Electronics}, {\bf 33}, 231 (2003)







\end{thebibliography}
\end{document}